\documentclass[aps,showpacs,showkeys,preprint,superscriptaddress]{revtex4}
\usepackage[caption = false]{subfig}
\usepackage{graphicx}
\usepackage{float}
\usepackage{hyperref}
\usepackage[left]{lineno}
\usepackage{blindtext}
\usepackage{bm}
\usepackage{amsmath}
\usepackage{txfonts}
\usepackage{epstopdf}
\usepackage{natbib}
\begin{document}
\title{Phase control of Schwinger pair production by colliding laser pulses
}
\author{Chitradip Banerjee\footnote{Corresponding author.\\E-mail address: cbanerjee@rrcat.gov.in (C. Banerjee).}}
\author{Manoranjan P. Singh}
\affiliation{Theory and Simulations Lab, HRDS, Raja Ramanna Centre for Advanced
Technology,  Indore-452013, India}
\affiliation{Homi Bhaba National Institute, Training School Complex, Anushakti Nagar, Mumbai 400094, India}
\date{\today}
\author{Alexander M. Fedotov}
\affiliation{National Research Nuclear University MEPhI, Moscow 115409, Russia}
\affiliation{Laboratory of Quantum Theory of Intense Fields, Tomsk State University, Lenin Prospekt 36, 634050, Tomsk, Russia}
\begin{abstract}
We study the Schwinger electron-positron pair production by a strong electromagnetic field of two colliding e-polarized laser pulses with a relative phase shift $\Psi$. The spatio-temporal distribution of created pairs is very sensitive to this phase shift and to polarization of the pulses. We study this dependence in detail and demonstrate how it can be explained in terms of the underlying invariant field structure of the counterpropagating focused pulses. 
\end{abstract}

\maketitle
\section{Introduction}
The $e^+e^-$ pair production from vacuum by strong electromagnetic (EM) field is a fundamental prediction of quantum electrodynamics (QED) \cite{berestetskii2012quantum,dirac1930theory,viritus,RevModPhys.84.1177DiPiazza}. Although the process was foreseen theoretically several decades ago \cite{Sauter,Heisenberg,PhysRev.82.664,VolkovDM,NarozhnyiNikishov1970,Nikishov1970346,PhysRevD.2.1191Itzykson}, its experimental verification is still missing because of the unavailability of an electric field strength comparable to the Schwinger limit $E_S= 1.32\times 10^{18}\:\textnormal{V/m}$. Since the probability $P_{e^+e^-}$ of pair production from vacuum by a strong electric field of strength $E_{peak}$ is proportional to $\exp(-\pi E_S/E_{peak})$, the process is exponentially suppressed for $E_{peak}\ll E_S$. Exploration of graphene \cite{SchwingerGraphene2008,PhysRevBGrapheneSchw2015,PhysRevBGrapheneSchw2016}, an isomorphic  $2D$-system in condensed matter physics, was proposed as a workaround to validate the Schwinger mechanism experimentally.
 
The available electric field strength for the present-day laser systems is of the order of $E_{peak}\sim 10^{13}-10^{14}\:\textnormal{V/m}$ \cite{RevModPhys.78.309Mourou2006,Yanovsky:08}, considerably below the critical field limit $E_S$. However, recent advances in technologies of  ultrashort and ultraintense laser pulse generation \cite{PhysRevSTAB.5.031301} raise the hopes that in a foreseeable future the available laser intensity may closer approach the elusive threshold of pair production. Moreover, such nonlinear QED effects as $e^+e^-$-pair photoproduction by a hard photon \cite{DbrukePhysRevLett.79.1626,PhysRevLett.101.200403Bell,PhysRevLett.106.035001Nerush} and the nonlinear (multiphoton) Compton scattering have been already observed experimentally at laser intensity $I = 10^{22}\textnormal{W/m}^2$ \cite{CbulaPhysRevLett.76.3116}. These developments renewed interest in theoretical studies of pair production by intense optical lasers.  

On the other hand, using the realistic focused field models, e.g. a weakly focused field in paraxial approximation \cite{Narozhny2000}, tightly focused field models \cite{fedotov2009electron,PhysRevSTAB_salamin}, and the optimally focused field model of e-dipole pulses \cite{PhysRevA2012Su,PhysRevLett.109.253202Su}, it was demonstrated that pair production can take place even at intensities substantially lower than the critical intensity $I_S = \frac{c}{4\pi}E_S^2$ (here $c$ denotes the light velocity in vacuum). 
 Superposition of laser pulses in a counterpropagating configuration has been shown  \cite{bula,PhysRevLett.104.220404,PhysRevD.82.105026Hebenstreit,PhysRevE.71.016404,PhysRevE.69.036408,PhysRevD.91.125026Wollert,PhysRevA.86.2012Gonoskov} to lower the threshold value of the required field strength considerably \cite{bula}. Such beam configurations were extensively used to study various aspects of pair production, including the dynamics of post production  \cite{PhysRevDSPKim2002,PhysRevDSPKim2007,PhysRevD.73SPKim2006,Dunne2009,PhysRevDSPKim2008,PhysRevDSPKim2009,Hebenstreit2014189,PhysRevD.80Chervyakov2009,PhysRevDHGies2005,PhysRevDHKleinert2008,PhysRevA.86.2012Gonoskov,PhysRevA2012Su,AbdukerimCarrier,PhysRevLett_CEP2009,PhysRevLettDumluStokes,PhysRevLettGiesHolger,PhysRevD.82Dumlu,PhysRevD.91.125026Wollert,PhysRevD.88.045028Kohlfurst,PhysRevE.71.016404,Banerjee2017,ChitradipCEP}.  

To the best of our knowledge, previous studies of pair production by counterpropagating laser pulses in realistic 3D setup almost never considered the effect of phase shift between the colliding pulses. However, it has been shown recently in Ref.~\cite{ChitradipCEP} that for a focused linearly polarized  standing wave the invariant electric field  distribution, and hence also pair production, are sensitive to the carrier envelope phase (CEP) $\tilde{\varphi}$. Here we study the spatio-temporal distribution of $e^+e^-$ pairs created via the Schwinger mechanism at a focal region of the colliding laser pulses described by the Narozhny-Fofanov model \cite{Narozhny2000}, assuming that the  pulses are in addition mutually phase shifted.  As we demonstrate, the phase shift $\Psi$ considerably affects the longitudinal spatial (here, z-) coordinate and time distributions of the resulting EM field, especially for ultrashort (few cycle) laser pulses. Furthermore, we study the dependence of the invariant field structure and of the distribution of the created pairs on polarization, relative sense of rotation (for circular polarization), and CEP of the counterpropagating pulses. 

The paper is organized as follows: in Sec.~\ref{theory} we briefly discuss the basic theory: the Schwinger formula for average pair production and the structure of the invariant electric and magnetic fields of the coherently superposed counterpropagating focused laser pulses. Next we present the differential pair production rates in spatiotemporal coordinates and explanation of their features in terms of the invariant electric field distribution in Sec.~\ref{Results}, finally concluding in Sec.~\ref{conclusion}. A technically useful  simplification of the envelope of counterpropagating pulses is discussed in Appendix~\ref{Appendix}.

\section{Theory}\label{theory}
\subsection{Methodology for calculating $e^+e^-$ pair production}\label{Schwinger_formula}
Assuming the validity of a locally constant field approximation, the average number of created pairs per unit time and volume can be calculated using the Nikishov formula \cite{PhysRevLettBulanov,nikishov1970pair}:
 \begin{equation}
  \label{w_e_average}
  w_{e^- e^+} = \frac{d^2N_{e^- e^+}}{dVdt} = \frac{e^2 E_S^2}{4 \pi^2 \hbar^2 c}  \epsilon \eta \coth\left(\frac{\pi \eta}{\epsilon}\right)\exp\left(-\frac{\pi}{\epsilon}\right),
  \end{equation}
where $e$ is the magnitude of the electron charge, and $\epsilon,\,\eta = \sqrt{\sqrt{{\mathcal{F}}^2+{\mathcal{G}}^2}\pm\mathcal{F}}$ [with $\mathcal{F}= \frac{1}{2}\Big(\textbf{E}^2-\textbf{H}^2\Big)$ and $\mathcal{G}=\textbf{E}\cdot\textbf{H}$] are the normalized (by $E_S$) invariant electric and magnetic field strengths, i.e. the magnitudes of the electric and magnetic field strengths in a reference frame where they are locally either zero or parallel.
Eq.~(\ref{w_e_average}) is valid in a locally constant field approximation based on an assumption that the characteristic length and time scales of the $e^+e^-$ pair production  process (the Compton length $\hbar/{m_ec}$ and time $\hbar/{m_ec^2}$ scales) are much smaller than the carrier wavelength ($\lambda \approx 1\mu\textnormal{m}$) and the period ($\lambda/c\approx 3\textnormal{fs}$) of the laser field, respectively \cite{PhysRevLettBulanov}. 
In particular, pair production is negligible if $\epsilon$ is small or vanishing in a focal region, while in the opposite case of nearly vanishing $\eta$ Eq.~(\ref{w_e_average}) reduces to
 \begin{equation}
   \label{w_e_average1}
   w_{e^- e^+} = \frac{d^2N_{e^- e^+}}{dVdt} \approx \frac{e^2 E_S^2}{4 \pi^3 \hbar^2 c}  {\epsilon}^2\exp\left(-\frac{\pi}{\epsilon}\right).
   \end{equation} 
As we will see later on, these special cases are realized for the magnetic and electric regimes in the focal region for collision of linearly polarized laser pulses. Hence we use Eq.~(\ref{w_e_average1}) for presenting the numerical results of differential particle production rates in spatiotemporal coordinates for linearly polarized laser pulses and Eq.~(\ref{w_e_average}) otherwise. To obtain a temporal particle distribution we integrate $w_{e^+e^-}$ over the spatial coordinates, and to obtain the longitudinal spatial distribution of particle production we integrate the production rate $w_{e^+e^-}$ over the transverse spatial coordinates and time. The actual distribution of the invariant fields in a focal region of colliding pulses strongly depends on their polarization and is discussed below. In all numerical calculations, we use the exact expressions for the EM fields and assume for definiteness the amplitude $E_0 = 0.0565$, carrier wavelength $\lambda = 1\mu m$, focusing parameter $\Delta = 0.1$, and pulse duration $\tau = 10 fs$ for each of the counterpropagating pulses. However, to easier interpret the results, in the rest of the section we also derive the approximate analytical expressions for field invariants near the focus.

\subsection{Invariant fields (linear polarization)}\label{field_expression}

Let us start with a field configuration of linearly polarized counterpropagating focused laser pulses based on the Narozhny-Fofanov field model \cite{Narozhny2000}. We assume the normalized (by $E_S$) electric fields of the pulses propagating in a forward ($+z$) and backward ($-z$) directions of the form
 \cite{bula}
 \begin{equation}\label{E_f_single}
 \textbf{E}_f = iE_0e^{-i\omega(t-z/c)-i\tilde{\varphi}}g\Bigg[\hat{\textbf{e}}_x(F_1-F_2\cos{2\phi})-\hat{\textbf{e}}_yF_2\sin{2\phi}\Bigg],
 \end{equation}
 and
 \begin{equation}\label{E_b_single}
 \textbf{E}_b = iE_0e^{-i\omega(t+z/c)-i\tilde{\varphi}-i\Psi}g\Bigg[\hat{\textbf{e}}_x(F_1^*-F_2^*\cos{2\phi})-\hat{\textbf{e}}_yF_2^*\sin{2\phi}\Bigg],
 \end{equation}
respectively, where $E_0$ is the normalized (by $E_S$) peak electric field strength of the laser pulse; $\omega= 2\pi c/\lambda$ is the central frequency of the pulse; $\lambda$ is the laser carrier wavelength; $F_1,\: F_2$ are the Gaussian-like functions of the form \cite{Narozhny2000} 
  \begin{equation}\label{Form_fuctions}
  \begin{split}
  F_1 = \frac1{(1+2i\chi)^{2}} \left( 1-\frac{\xi^2}{1+2i\chi} \right) \exp\left(-\frac{\xi^2}{1+2i\chi}\right),\quad 
  F_2 = -\frac{\xi^2}{(1+2i\chi)^3} \exp\left(-\frac{\xi^2}{1+2i\chi}\right),\nonumber
 \end{split}
 \end{equation}
$F_1^*$ and $F_2^*$ are their complex conjugates; $\xi = \rho/R$ is the normalized radial variable with $\rho = \sqrt{x^2+y^2}$ at the transverse Cartesian spatial coordinates $x$, $y$; $R$ is the focal radius; $\phi = \arctan(y/x)$ is the azimuthal angle; $\chi = z/L$ is the normalized longitudinal coordinate with $L = R/\Delta$ being the Rayleigh length for a focusing aperture parameter $\Delta = c/{\omega R}$. The envelope function $g$ accounts for temporal finiteness of the laser pulses. In this paper we take $g = \exp(-4t^2/\tau^2-4z^2/{c^2\tau^2})$ \cite{Banerjee2017,ChitradipCEP} (a detailed explanation of our method of introducing $g$ is given in Appendix~\ref{Appendix}). Finally, $\tilde{\varphi}$ and $\Psi$ are CEP and the phase shift of the backward propagating pulse, respectively.
 
The corresponding expressions for the normalized magnetic field of the forward and backward propagating pulses are \cite{bula} 
 \begin{equation}\label{H_f_single}
 \textbf{H}_f = iE_0e^{-i\omega(t-z/c)-i\tilde{\varphi}}g\Bigg[\left(1-i\Delta^2\frac{\partial}{\partial\chi}\right)\Big\{\hat{\textbf{e}}_xF_2\sin{2\phi}-\hat{\textbf{e}}_y(F_1-F_2\cos{2\phi})\Big\}+2i\Delta\sin{\phi}\frac{\partial F_1}{\partial\xi}\hat{\textbf{e}}_z\Bigg],
 \end{equation}
 and 
 \begin{equation}\label{H_b_single}
 \textbf{H}_b = -iE_0e^{-i\omega(t+z/c)-i\tilde{\varphi}-i\Psi}g\Bigg[\left(1+i\Delta^2\frac{\partial}{\partial\chi}\right)\Big\{\hat{\textbf{e}}_xF_2^*\sin{2\phi}-\hat{\textbf{e}}_y(F_1^*-F_2^*\cos{2\phi})\Big\}+2i\Delta\sin{\phi}\frac{\partial F_1^*}{\partial\xi}\hat{\textbf{e}}_z\Bigg].
 \end{equation}

Following the procedure of Ref.~\cite{ChitradipCEP}, the expressions for the Lorentz invariants of the resultant EM field $\textbf{E}=\textbf{E}_f+\textbf{E}_b$, $\textbf{H}=\textbf{H}_f+\textbf{H}_b$ of the superposed counterpropagating pulses can be derived,
 \begin{equation}\label{F_mu-1_lin}
        \begin{split}
        \mathcal{F} = \frac{1}{2}\Big({Re\textbf{E}}^2-{Re\textbf{H}}^2\Big)\approx \frac{2E_0^2g^2e^{-\frac{2\xi^2}{{1+4\chi^2}}}}{(1+4\chi^2)^{2}}\Bigg[\sin^2(\omega t+\tilde{\varphi}+\Psi/2)-\sin^2(\omega z/c+\Psi/2)\Bigg],
        \end{split}
\end{equation}
and
\begin{equation}\label{G_mu-1_lin}
\begin{split}
\mathcal{G} = Re\textbf{E}\cdot Re\textbf{H}^e\approx \frac{2E_0^2g^2\xi^2e^{-\frac{2\xi^2}{{1+4\chi^2}}}}{(1+4\chi^2)^{5/2}}\sin{(2\phi)}\sin{[2(\omega t+\tilde{\varphi}+\Psi/2)]}\sin{[2(\omega z/c +\Psi/2)]},
\end{split}
\end{equation}
where we retain only the leading order terms in $\Delta$, $\xi$, and $\chi$, as justified in the focal region in a weak focusing limit. Since $\mathcal{G}=\mathcal{O}\left(\xi^2\right)$ is negligibly small there, one of the invariant fields (depending on the sign of $\mathcal{F}$) is vanishingly small. For $\mathcal{F} > 0$ we have so-called electric regime \cite{ElectricregimeFedotov}
\begin{equation}\label{epsilon_mu-1_lin1}
         \begin{split}
         \epsilon_{elec} \approx \frac{2E_0ge^{-\frac{\xi^2}{{1+4\chi^2}}}}{(1+4\chi^2)}\Bigg[\sin^2(\omega t+\tilde{\varphi}+\Psi/2)-\sin^2(\omega z/c+\Psi/2)\Bigg]^{1/2}, \:\: \:\textnormal{and}\:\: \eta_{elec} \approx 0,
         \end{split}
         \end{equation}
whereas, for a magnetic regime $\mathcal{F} < 0$ \cite{ElectricregimeFedotov}
\begin{equation}\label{epsilon_mu-1_lin2}
         \begin{split}
         \epsilon_{mag} \approx 0, \:\: \:\textnormal{and}\:\:\eta_{mag} \approx  \frac{2E_0ge^{-\frac{\xi^2}{{1+4\chi^2}}}}{(1+4\chi^2)}\Bigg[\sin^2(\omega z/c+\Psi/2)-\sin^2(\omega t+\tilde{\varphi}+\Psi/2)\Bigg]^{1/2}. 
         \end{split}
         \end{equation}
Clearly, pairs are created solely during an electric regime, and the phases $\Psi$ and $\tilde{\varphi}$ control toggling between the electric and magnetic regimes at given point and time, thereby controlling also the pair production. As is seen from the obtained approximate expressions (and in fact is also true for the exact ones), it is enough to restrict phases by $0\le\tilde{\varphi}<\pi$ and $0\le\Psi<2\pi$.

\begin{figure}[h]
\begin{center}
          \subfloat[$~\tilde{\varphi} = 0,~\Psi = 0$]{ \includegraphics[scale = .33]{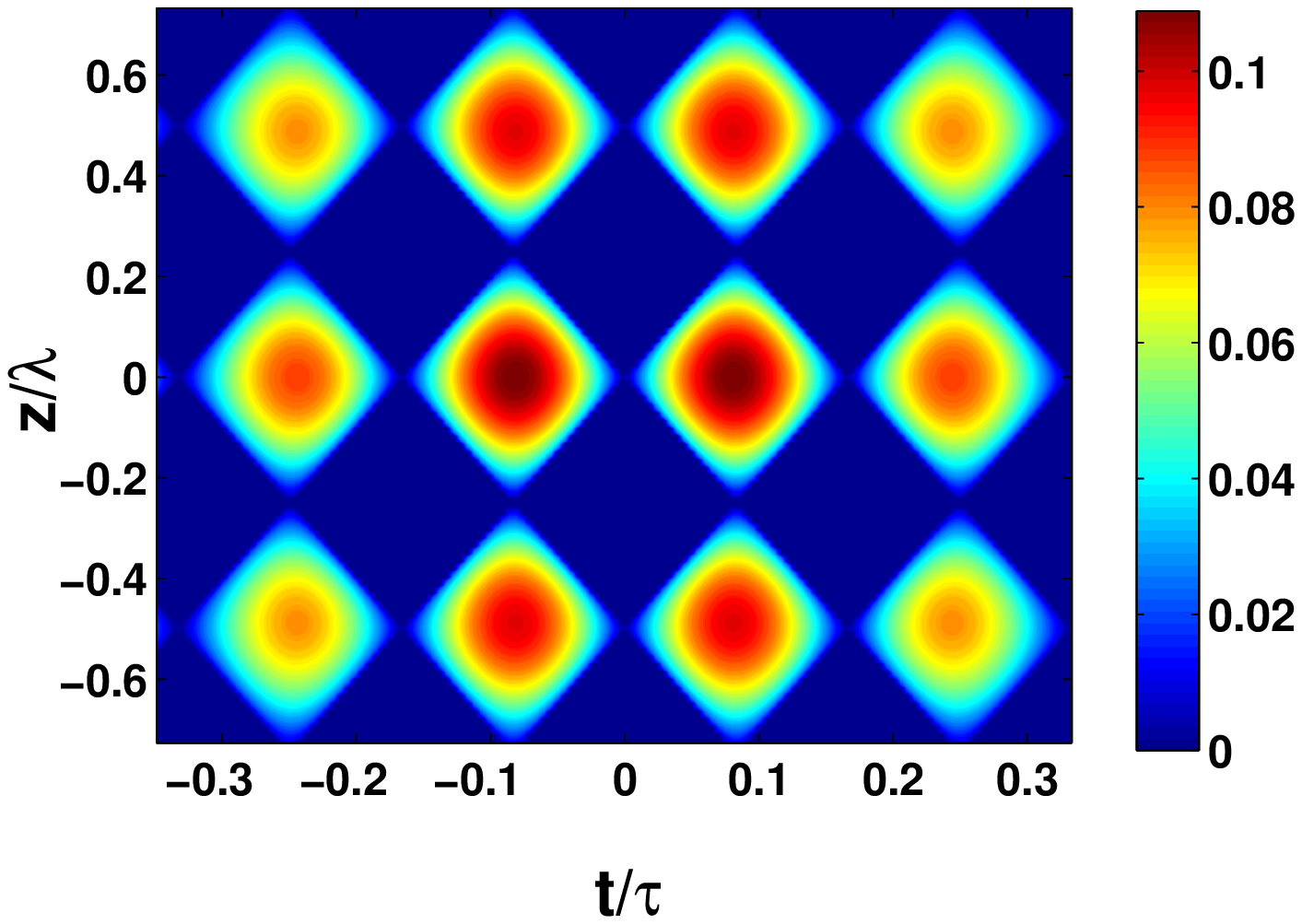}}
          \subfloat[$~\tilde{\varphi} = 0,~\Psi = \pi/2$]{ \includegraphics[scale = .33]{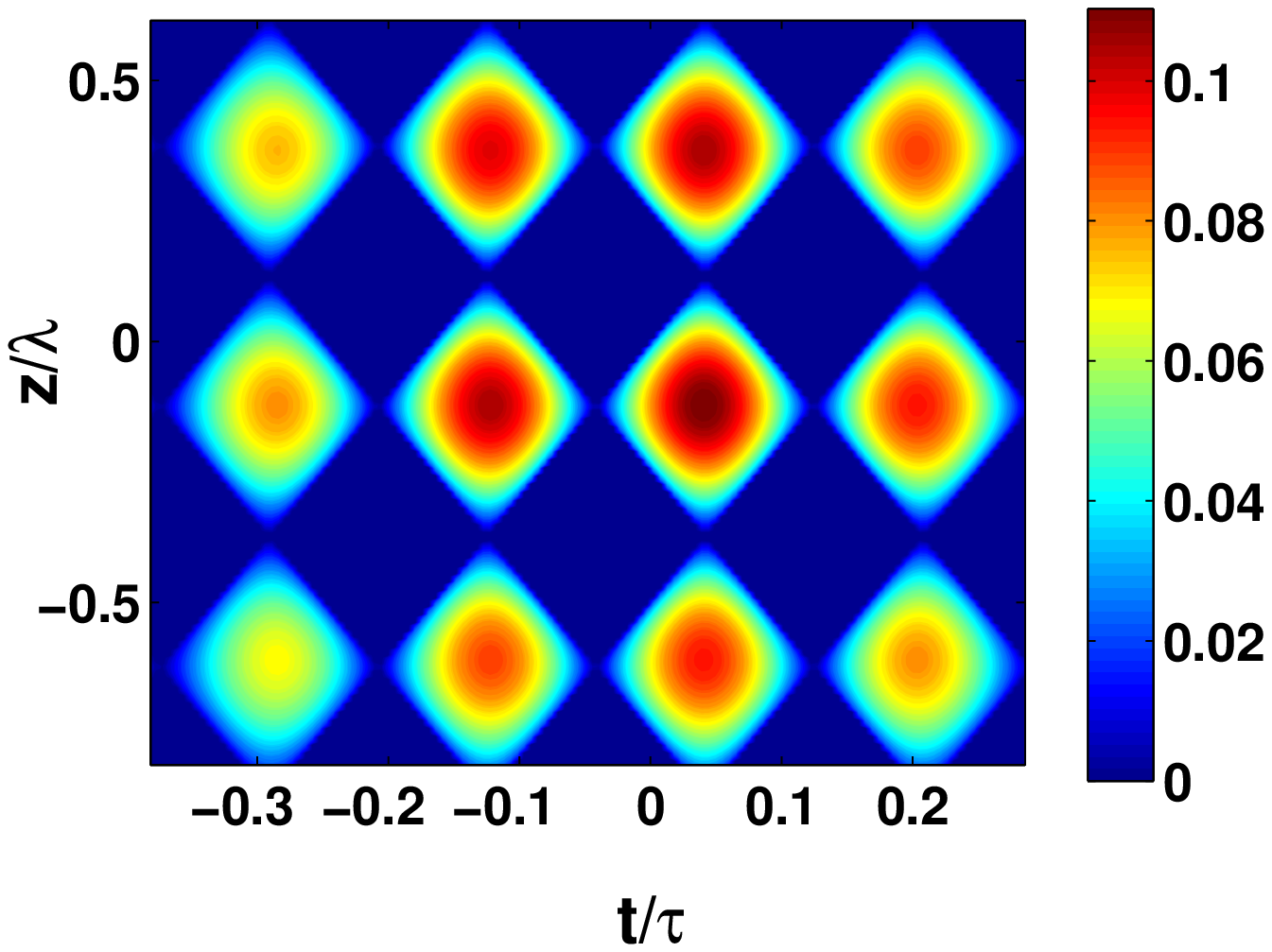} }
          \subfloat[$~\tilde{\varphi} = 0,~\Psi = \pi$]{ \includegraphics[scale = .33]{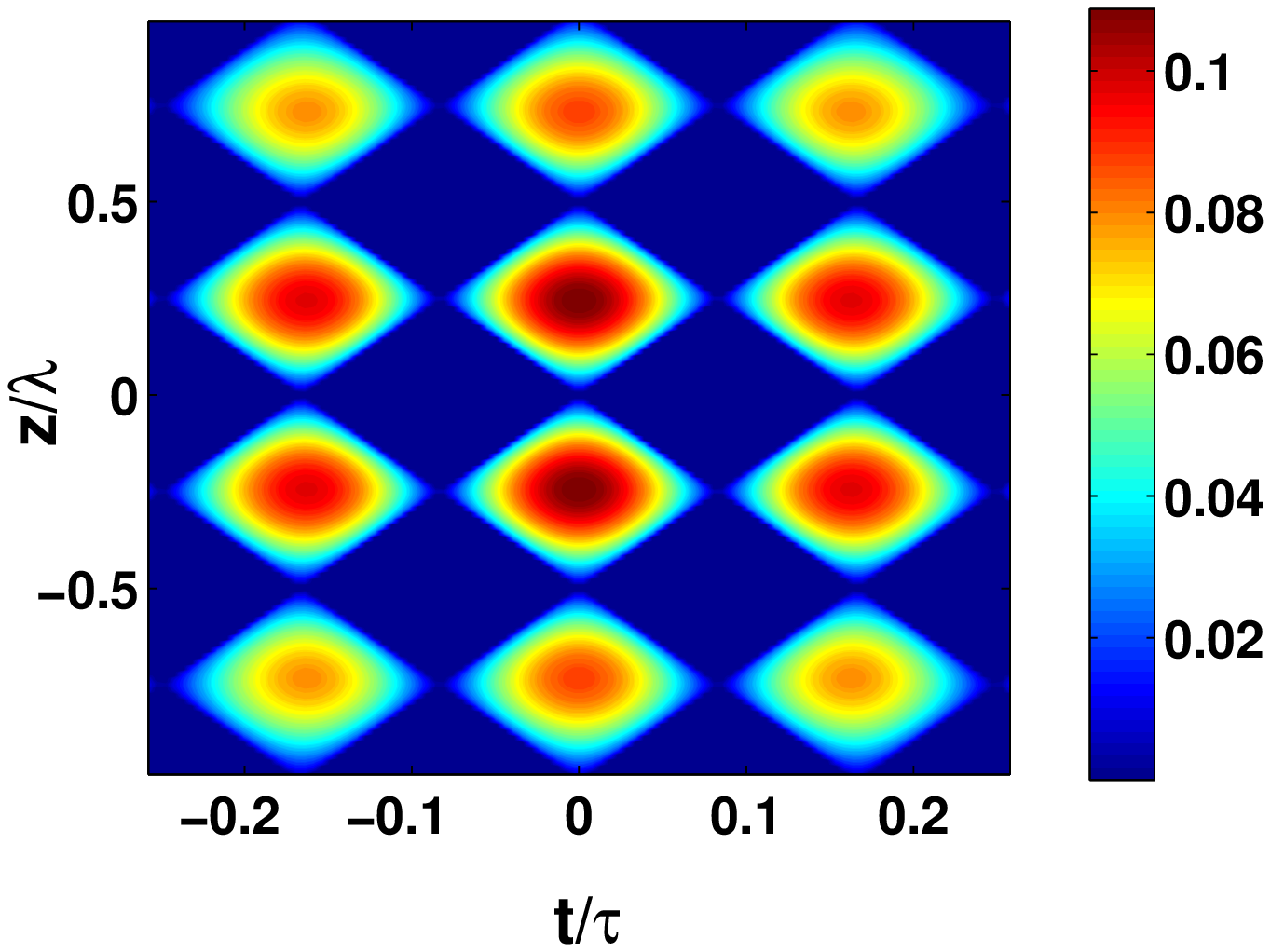}}\\
           \subfloat[$~\tilde{\varphi} = \pi/2,~\Psi = 0$]{\includegraphics[scale = .33]{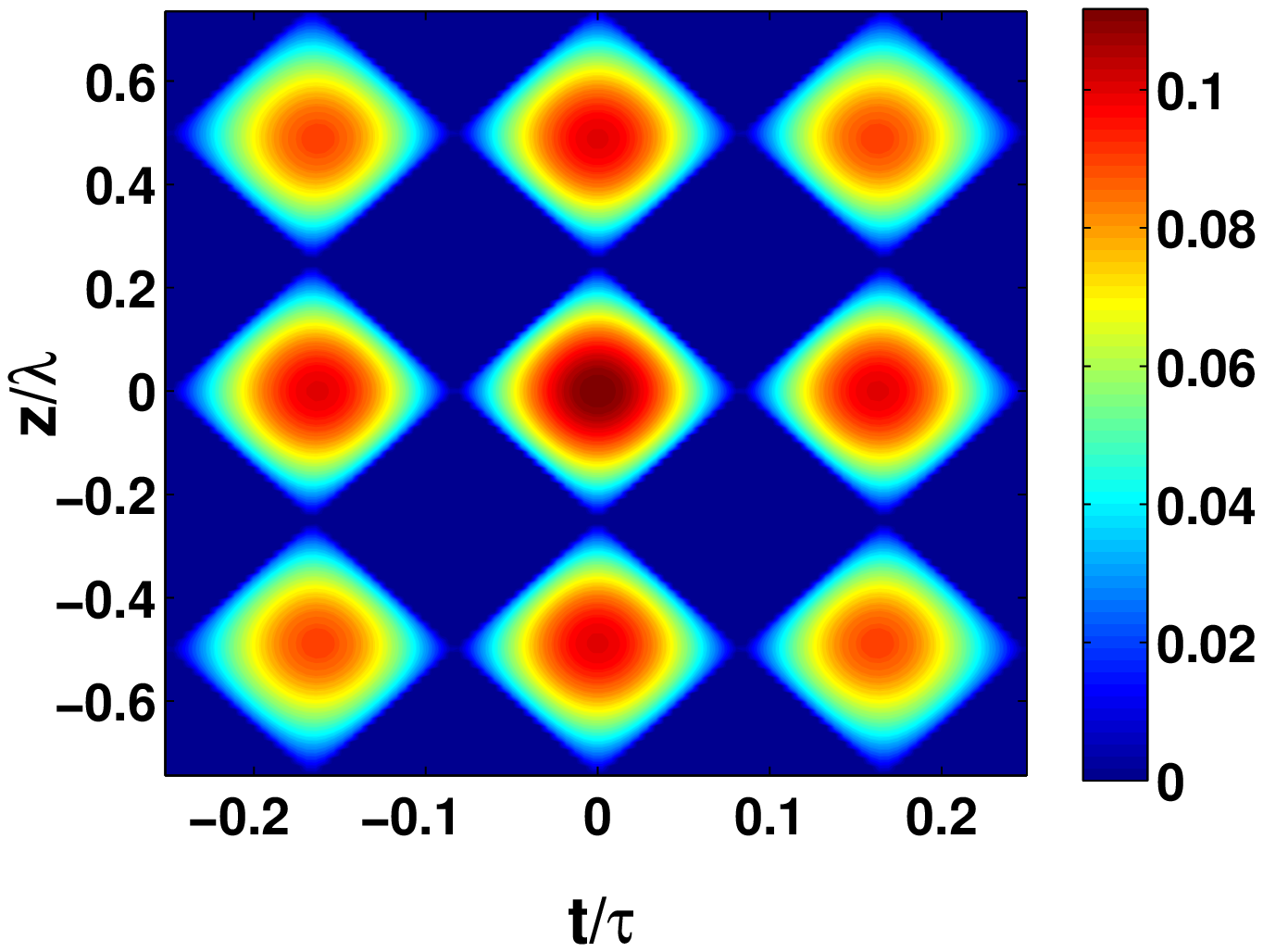} }
           \subfloat[$~\tilde{\varphi} = \pi/2,~\Psi = \pi/2$]{\includegraphics[scale = .33]{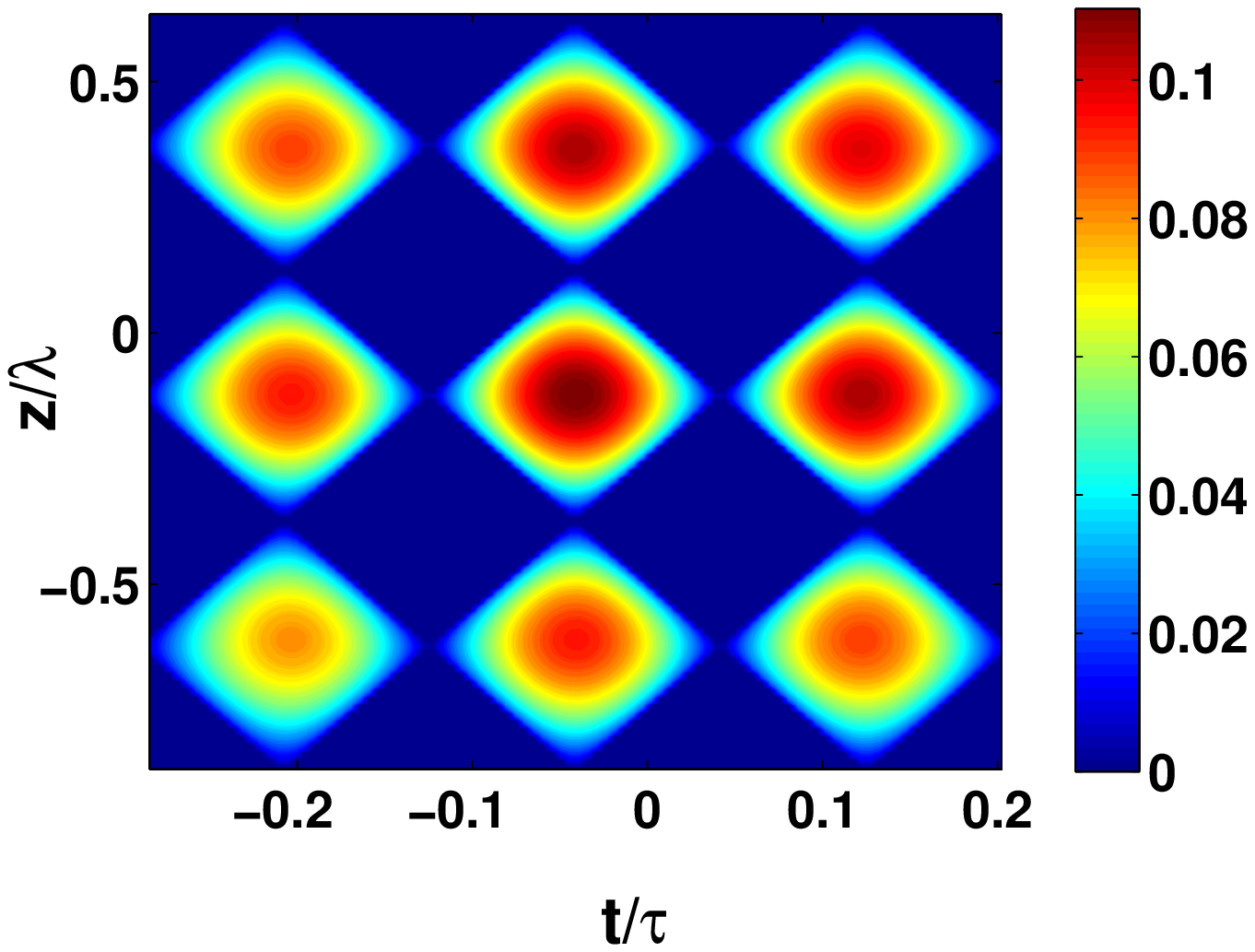}}
           \subfloat[$~\tilde{\varphi} = \pi/2,~\Psi = \pi$]{\includegraphics[scale = .33]{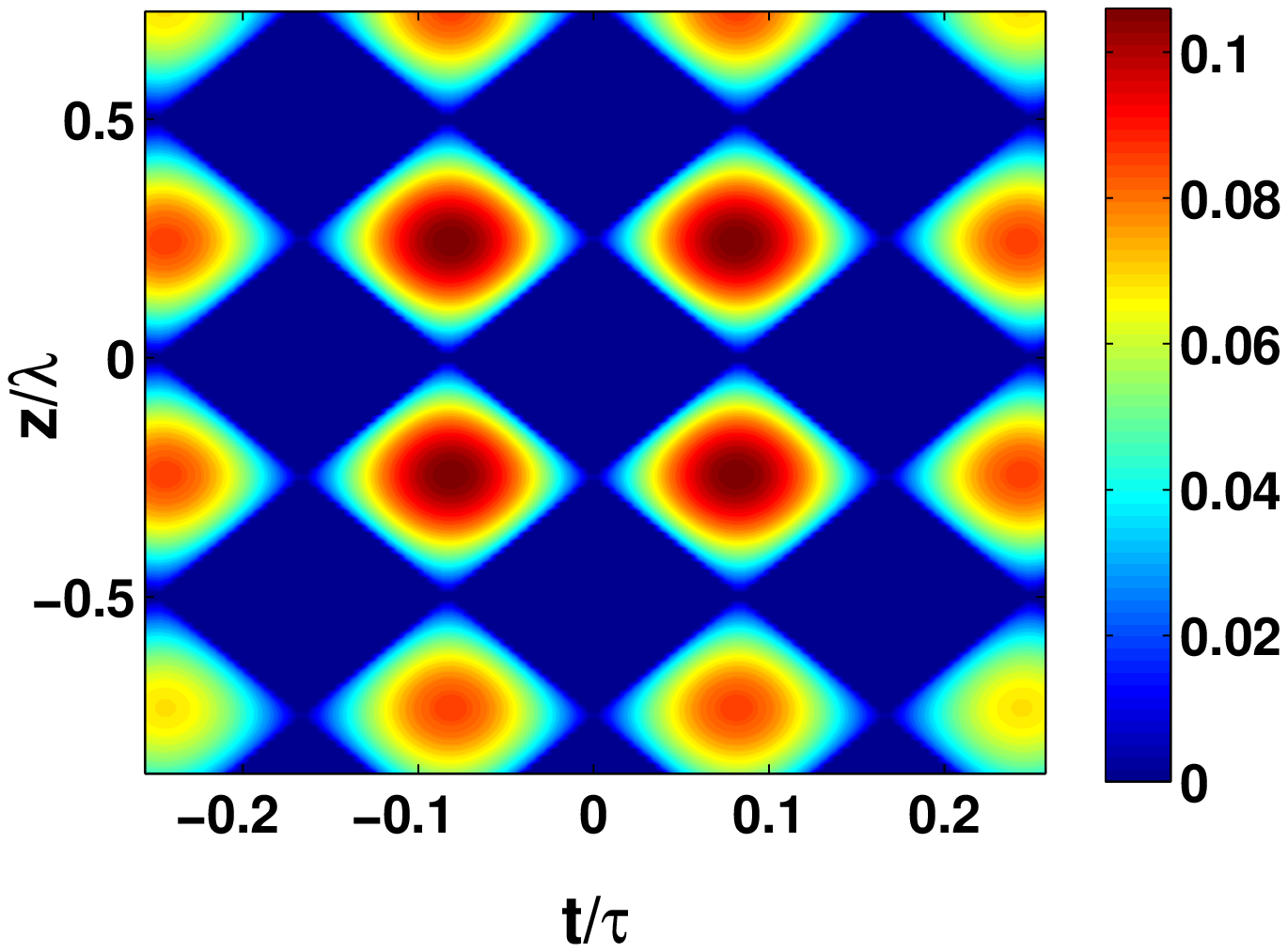}}
\caption{Spatiotemporal distributions of the invariant electric field $\epsilon (x=0,y=0,z,t)$ for linearly e-polarized Gaussian laser pulses colliding with different values of the phases $\tilde{\varphi}$ and $\Psi$. The laser parameters are $E_0 = 0.0565$, $\Delta = 0.1$, $\tau = 10fs$, and $\lambda = 1\mu m$.}
\label{epsilon_contour_lin}
\end{center}
\end{figure}

Spatiotemporal distributions of the invariant field $\epsilon^e$ for few representative values $\Psi=0,\;\pi/2,\;\pi$ and $\tilde{\varphi}=0,\;\pi/2$ are presented in Fig.~\ref{epsilon_contour_lin}, where the rhombic structure corresponds to the aforementioned separation into the alternating electric (color) and magnetic (dark) regimes. It is clear from the figure, as well as from the above equations, that the maxima are shifted with respect to the origin $t=z=0$, and that their shift is determined by the phases. If the maxima are remote from the origin (which is at the center of the envelope) then their magnitudes are reduced, in this way they are also indirectly controlled by the phases.
The figure illustrates a variety of possible opportunities: the maxima can be located symmetrically about the origin, with either one [see Fig.~\ref{epsilon_contour_lin}(d)] or a gap [Figs.~\ref{epsilon_contour_lin}(a,c,f)] seating at the origin (in the latter case there can be either four or two maxima closest to it: if there are two then they can be lined up along either axis); or off centered [like in Figs.~\ref{epsilon_contour_lin}(b,e)] -- in the latter case the magnitudes of the maxima are lined up according to their distance from the origin. Qualitatively, the pair production rate behaves the same way [see Eq.~(\ref{w_e_average}) or (\ref{w_e_average1})], hence, as we will see below, this variety of opportunities precisely corresponds to peculiarities of distribution of created pairs that we observe in our calculations.

\subsection{Invariant fields (circular polarization)}

Now consider circularly polarized forward and backward propagating laser pulses with a relative phase difference $\Psi$. We follow closely the steps discussed in Ref.~\cite{ChitradipCEP,Banerjee2017}. For a forward propagating (along $+z$) laser pulse, the electric and magnetic fields are given by
 \begin{equation}
  \label{Electric_field_e_single_f_RL}
   \textbf{E}_f = iE_0e^{-i\omega(t-z/c)-i\tilde{\varphi}}g\Big[F_1(\hat{\textbf{e}}_x \pm i\hat{\textbf{e}}_y)-F_2e^{ \pm 2i\phi}(\hat{\textbf{e}}_x \mp i\hat{\textbf{e}}_y)\Big],
   \end{equation}
  and
  \begin{equation}\label{Magnetic_field_e_single_f_RL}
  \textbf{H}_f =  \pm E_0e^{-i\omega(t-z/c)-i\tilde{\varphi}}g\Bigg[\left(1-i\Delta^2\frac{\partial}{\partial\chi}\right)\Big[F_1(\hat{\textbf{e}}_x \pm i\hat{\textbf{e}}_y)+F_2e^{ \pm 2i\phi}(\hat{\textbf{e}}_x \mp i\hat{\textbf{e}}_y)\Big]+2i\Delta e^{ \pm i\phi}\frac{\partial F_1}{\partial\xi}\hat{\textbf{e}}_z\Bigg],
  \end{equation}
respectively. Here the signs correspond to the right ($+$)- and left ($-$)- handed rotation of the electric field vector with respect to propagation direction. For a backward propagating (along $-z$) laser pulse with a relative phase shift $\Psi$ the expressions for the electric and magnetic fields are given by
  \begin{equation}
   \label{Electric_field_e_single_b_RL}
    \textbf{E}_b = iE_0e^{-i\omega(t+z/c)-i\tilde{\varphi}-i\Psi}g\Big[F_1^*(\hat{\textbf{e}}_x \pm i\hat{\textbf{e}}_y)-F_2^*e^{\mp 2i\phi}(\hat{\textbf{e}}_x \mp i\hat{\textbf{e}}_y)\Big],
     \end{equation}
   and
    \begin{equation}\label{Magnetic_field_e_single_b_RL}
   \textbf{H}_b = \mp E_0e^{-i\omega(t+z/c)-i\tilde{\varphi}-i\Psi}g\Bigg[\left(1+i\Delta^2\frac{\partial}{\partial\chi}\right)\Bigg[F_1^*(\hat{\textbf{e}}_x \pm i\hat{\textbf{e}}_y)+F_2^*e^{ \mp 2i\phi}(\hat{\textbf{e}}_x \mp i\hat{\textbf{e}}_y)\Big]+2i\Delta e^{\mp i\phi}\frac{\partial F_1^*}{\partial\xi}\hat{\textbf{e}}_z\Bigg].
   \end{equation}
   
For a pair of counterpropagating circularly polarized pulses, we have two alternatives: either both pulses have the same polarization (for definiteness right handed, hereafter referred to as the RR configuration), or opposite polarizations (for definiteness we assume that the forward propagating pulse has the right handed polarization and the backward propagating one has the left handed polarization, hereafter referred to as the RL configuration).

\subsubsection{RR Configuration}
\begin{figure}[H]
  \begin{center}
  \includegraphics[width = 5in]{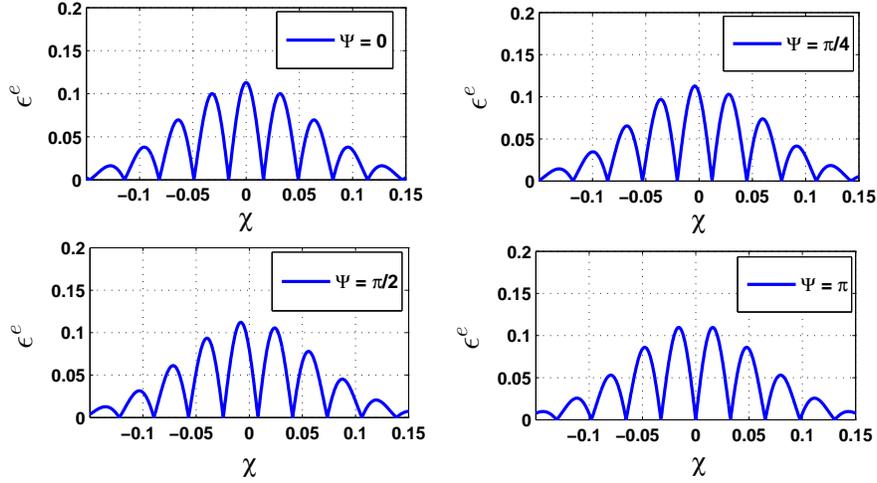} 
  \caption{Dependence of the invariant electric field $\epsilon$ at $\xi = \phi = 0$ on the longitudinal coordinate $\chi$ for counterpropagating circularly e-polarized focused Gaussian laser pulses in RR configuration with relative phases for $\Psi =  \:0,\:\pi/4,\:\pi/2,$ and $\pi$.  Laser parameters are the same as in Fig.~\ref{epsilon_contour_lin}.}
  \label{Epsilon_chi_RR_phase_cir}
  \end{center}
  \end{figure}

For the RR configuration the Lorentz invariants near the focus ($\xi,\chi\ll 1$) are given by 
\begin{equation}\label{FG_RR_cir}
 \begin{split}
 \mathcal{F}_{RR} \approx \frac{2E_0^2g^2e^{-\frac{2\xi^2}{{1+4\chi^2}}}}{(1+4\chi^2)^{2}}\cos{[2(\omega z/c+\Psi/2)]},\quad
 \mathcal{G}_{RR} \approx \frac{2E_0^2g^2e^{-\frac{2\xi^2}{1+4\chi^2}}}{(1+4\chi^2)^{2}}\sin{[2(\omega z/c+\Psi/2)]}, 
 \end{split}
 \end{equation}  
so that the invariant electric and magnetic fields are
   \begin{equation}\label{epsilon_eta_RR_cir}
    \begin{split}
    \epsilon_{RR}\approx \frac{2E_0ge^{-\frac{\xi^2}{{1+4\chi^2}}}}{(1+4\chi^2)}|\cos{(\omega z/c+\Psi/2)}|,\quad
        \eta_{RR} \approx \frac{2E_0ge^{-\frac{\xi^2}{{1+4\chi^2}}}}{(1+4\chi^2)}|\sin{(\omega z/c+\Psi/2)}|.
        \end{split}
        \end{equation}
One can see that, unlike the case of linearly polarized configuration, now their phases are time-independent and are solely controlled by single phase $\Psi$.  The overall smooth temporal dependence on a time scale $\tau$ remains only due to the pulse envelope function $g$.  The oscillatory dependence on longitudinal coordinate $\chi$ is shown in Fig.~\ref{Epsilon_chi_RR_phase_cir}. It is clear that as $\Psi$ is growing from zero, the highest central spike becomes off-centered, and eventually at $\Psi=\pi$ is replaced with two spikes of  equal height located symmetrically about the center.   

\subsubsection{RL Configuration}
\begin{figure}[H]
   \begin{center}
   \includegraphics[width = 5in]{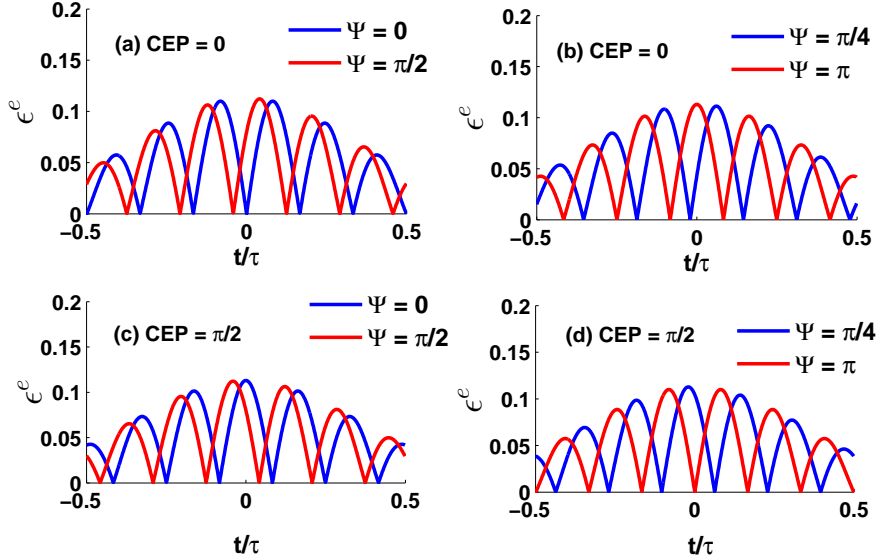} 
   \caption{Time evolution of the invariant electric field $\epsilon$ at $\xi = \phi =\chi= 0$ for counterpropagating circularly e-polarized focused Gaussian laser pulses in RL configuration with CEP $\tilde{\varphi} = 0,~\pi/2$ and with relative phase $\Psi =  \:0,\:\pi/4,\:\pi/2,$ and $\pi$. Laser parameters are the same as in Fig.~\ref{epsilon_contour_lin}.}
   \label{Epsilon_cir_RL_t_CEP_phase}
   \end{center}
   \end{figure}

Proceeding the same way for the RL configuration, we obtain the leading order expressions of the Lorentz invariants
\begin{equation}
\mathcal{F}_{RL} \approx -2E_0^2g^2\frac{e^{-\frac{2\xi^2}{1+4\chi^2}}}{(1+4\chi^2)^{2}}\cos{[2(\omega t+\tilde{\varphi}+\Psi/2)]}, \quad
\mathcal{G}_{RL} \approx 2E_0^2g^2\frac{e^{-\frac{2\xi^2}{1+4\chi^2}}}{(1+4\chi^2)^{2}}\sin{[2(\omega t+\tilde{\varphi}+\Psi/2)]},
\end{equation}
and the invariant electric and magnetic fields read as follows:
\begin{equation}\label{epsilon_eta_cir_CEP}
\begin{split}
\epsilon_{RL} \approx 2E_0g\frac{e^{-\frac{\xi^2}{1+4\chi^2}}}{(1+4\chi^2)}|\sin(\omega t+\tilde{\varphi}+\Psi/2)|,\quad
\eta_{RL} \approx 2E_0g\frac{e^{-\frac{\xi^2}{1+4\chi^2}}}{(1+4\chi^2)}|\cos(\omega t+\tilde{\varphi}+\Psi/2)|.
\end{split}
\end{equation}
In contrast to the RR case here their oscillations are purely temporal and depend on both $\Psi$ and $\tilde{\varphi}$, see Fig.~\ref{Epsilon_cir_RL_t_CEP_phase}. As it shows, variation of the phases, like in previous case, results in a shift of the main maximum from the center. Namely, it is located at the origin $t=0$ for $\tilde{\varphi} = 0$  at $\Psi = \pi$, whereas for $\tilde{\varphi} = \pi/2$ at $\Psi = 0$. As we will see in the next section, off-centering of the main maximum results in passing from unimodal to bimodal profile of created pairs.

\section{Pair production: results and discussion}\label{Results}

According to Eq.~(\ref{w_e_average}), the pair production rate depends exponentially and monotonously on the spatiotemporal distribution of the invariant electric field in the focal region, which in turn is controlled by the phase shifts $\Psi$ and $\tilde{\varphi}$. Hence, we present and discuss the results of calculation of differential particle production rate for various polarizations of the collided pulses and in dependence on the values of $\Psi$ and $\tilde{\varphi}$. The main goal is to demonstrate how its features can be natively understood in terms of the underlying invariant EM field spatiotemporal structure.

\subsection{Differential pair production rate (linear polarization)}\label{particleDist}
\begin{figure}[h]
\begin{center}
\includegraphics[scale = .32]{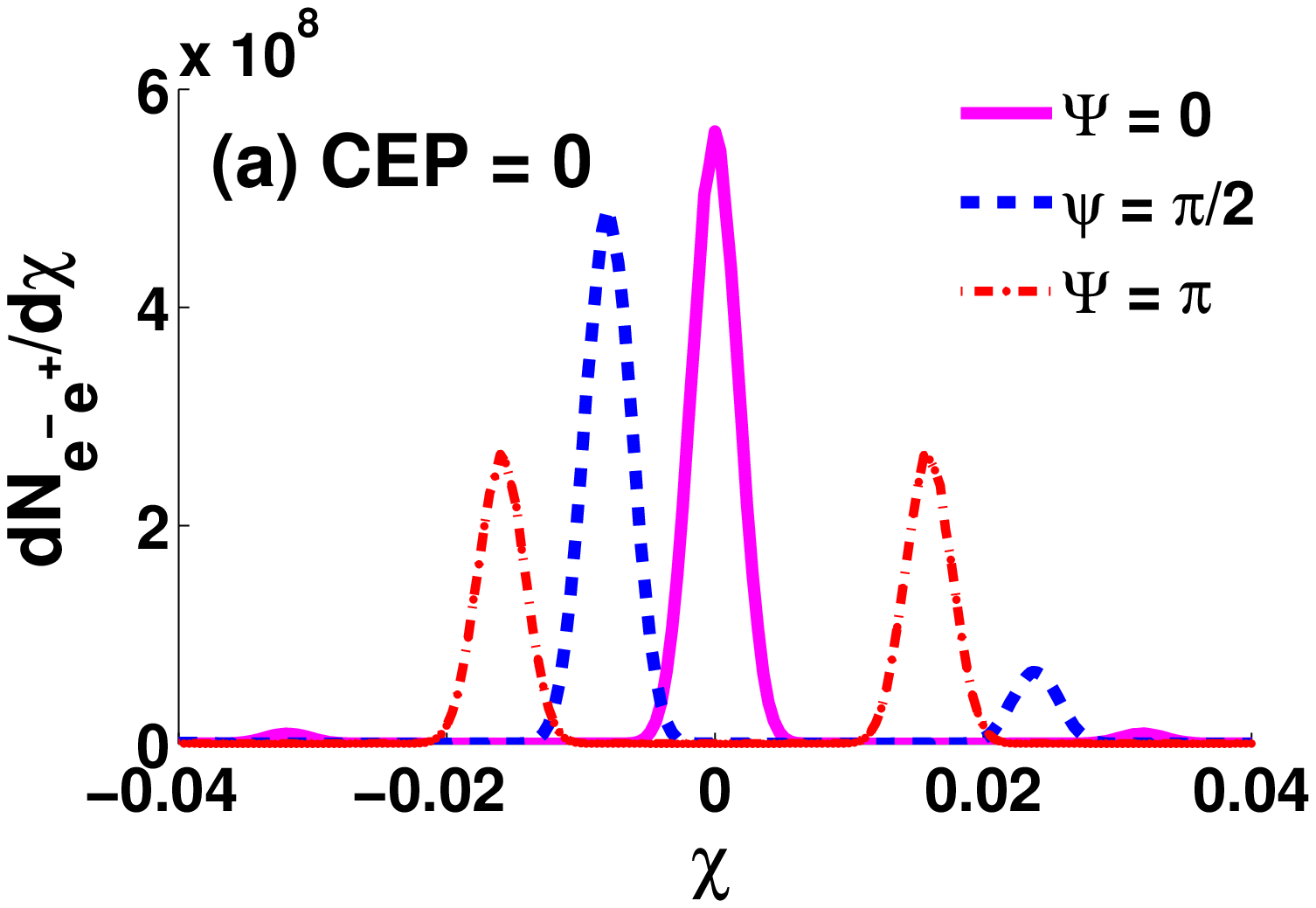}
\includegraphics[scale = .32]{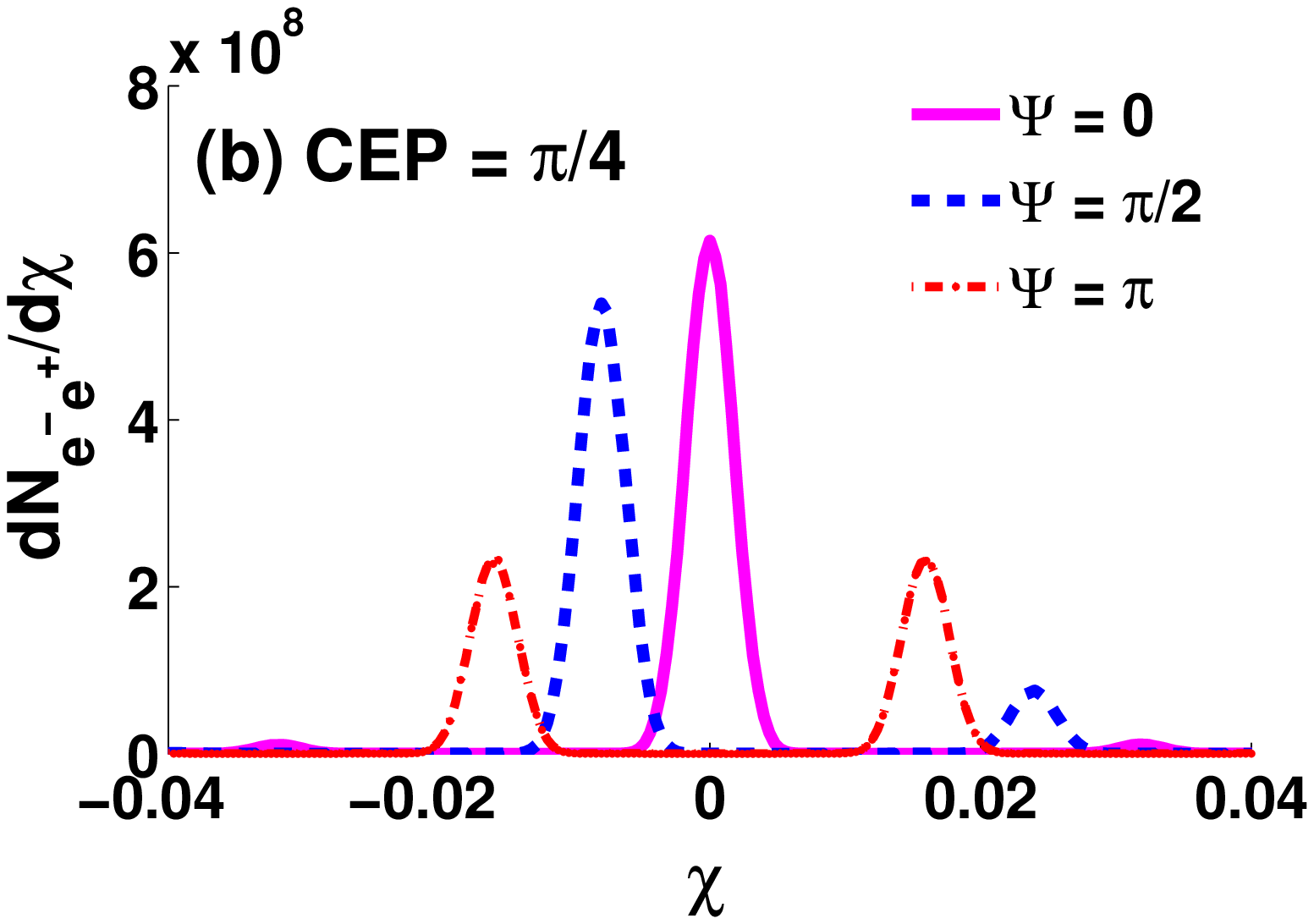} 
\includegraphics[scale = .32]{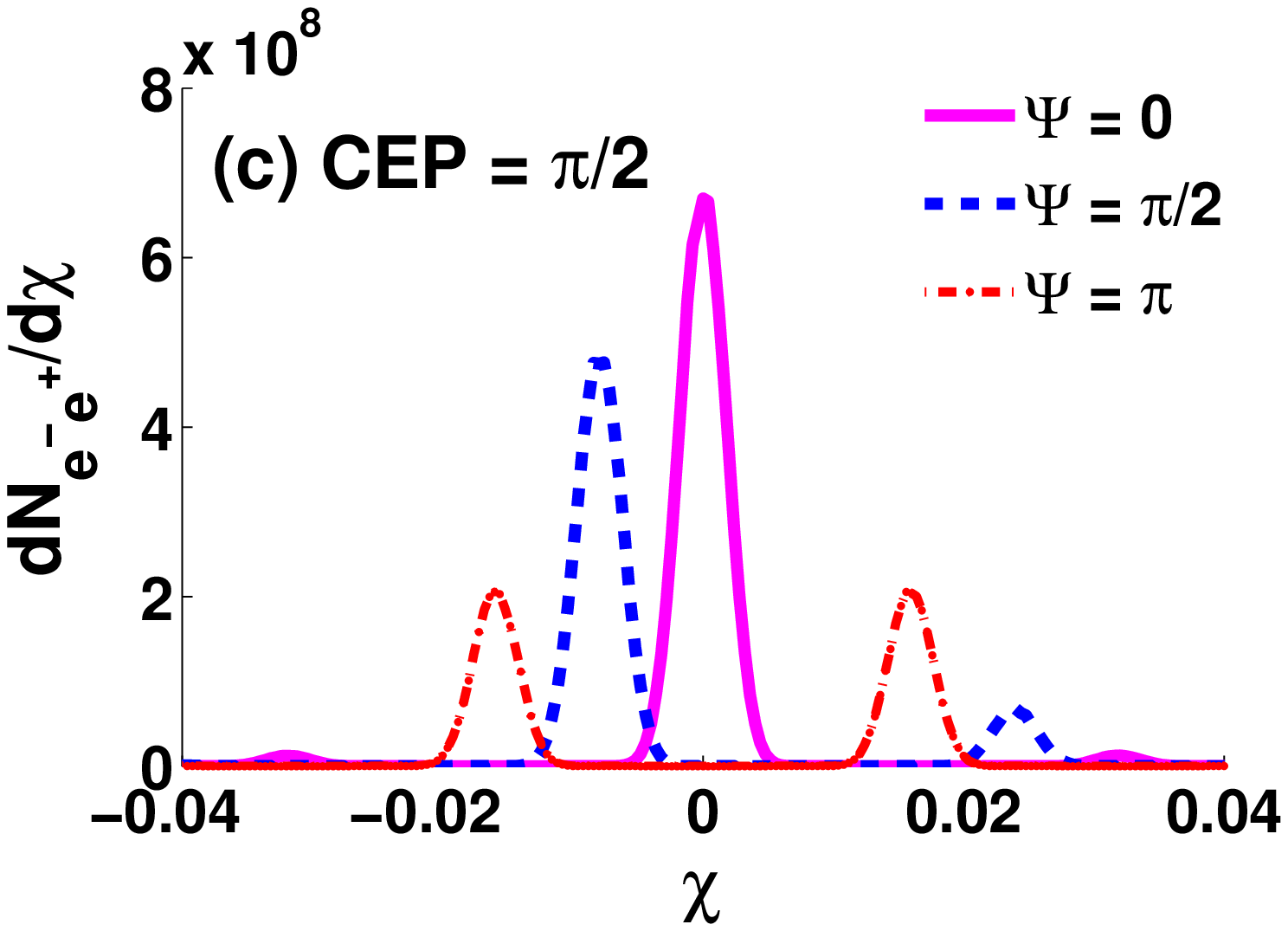}
\caption{Spatial distributions in longitudinal coordinate $\chi$ of particles created by colliding linearly polarized laser pulses with $\Psi=0,\;\pi/2,\;\pi$ and $\tilde{\varphi} = 0$, $\pi/4$, $\pi/2$. Laser parameters are the same as in Fig.~\ref{epsilon_contour_lin}.}
\label{dN_chi_CEP_psi}
\end{center}
\end{figure}

\begin{figure}[h]
\begin{center}
\includegraphics[scale = .32]{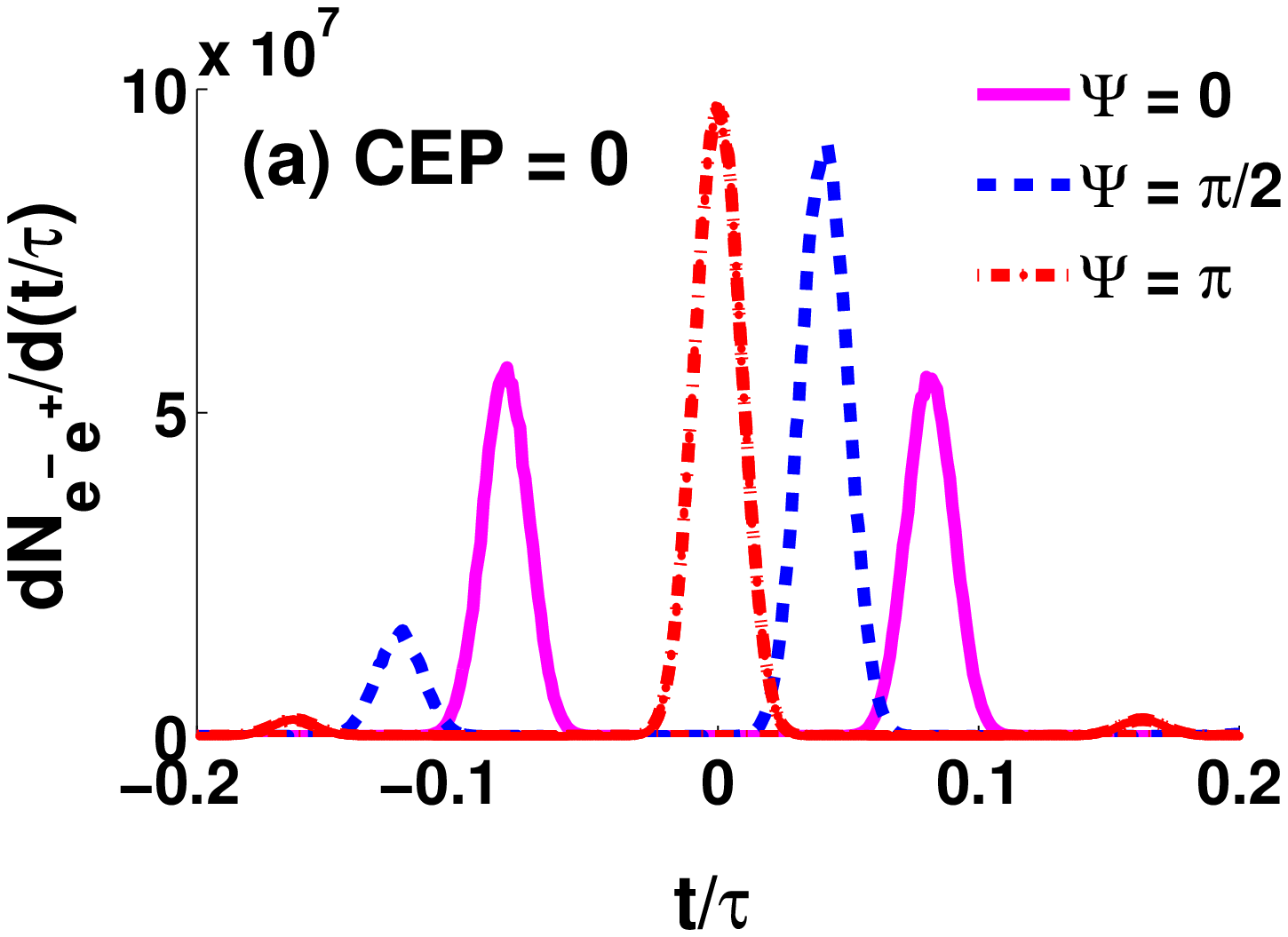}
\includegraphics[scale = .32]{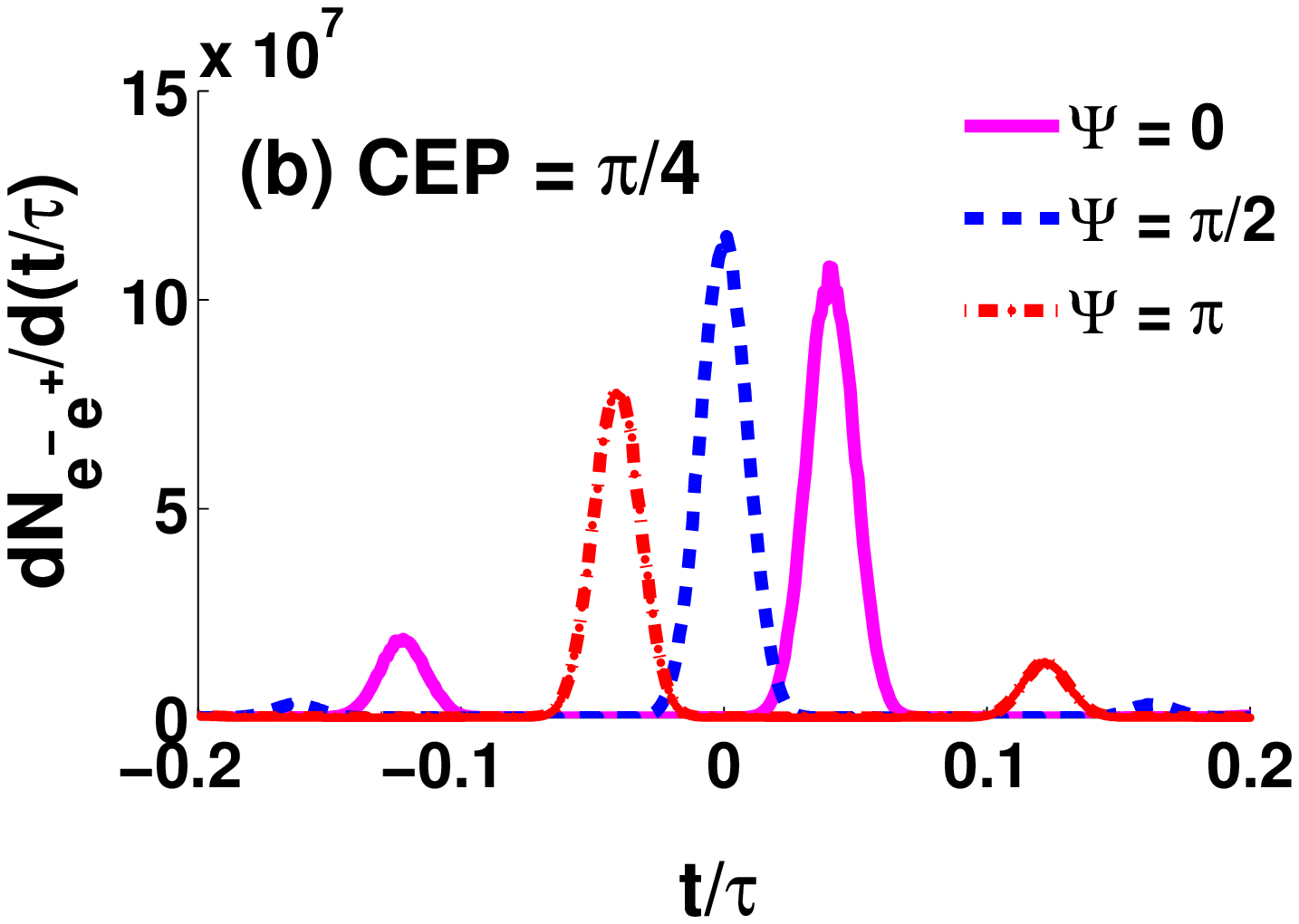}
\includegraphics[scale = .32]{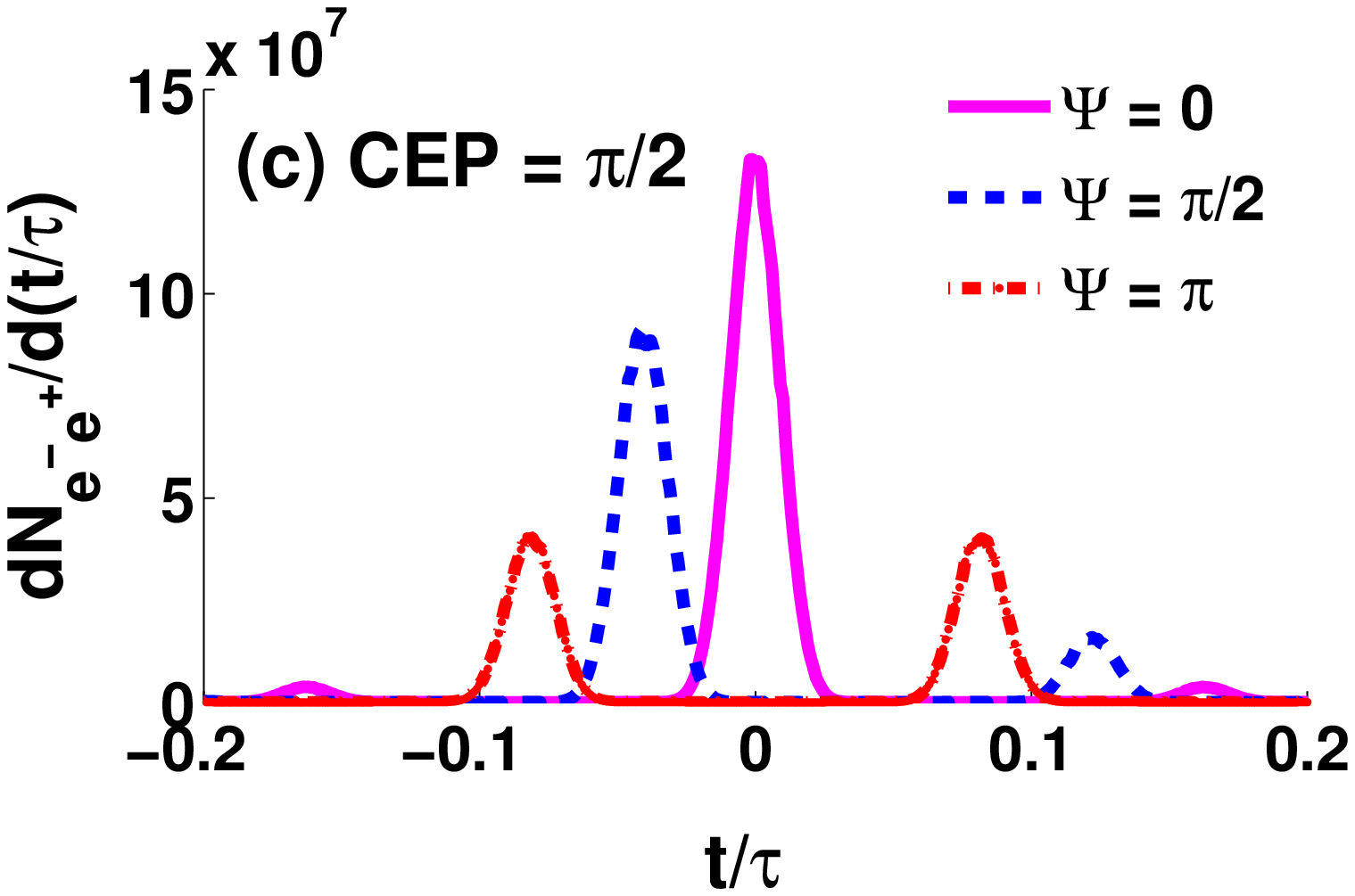}
\caption{Temporal distributions of particles created by colliding linearly polarized laser pulses with $\Psi=0,\;\pi/2,\;\pi$ and $\tilde{\varphi} = 0$, $\pi/4$, $\pi/2$. Laser parameters are the same as in Fig.~\ref{epsilon_contour_lin}.}
\label{dN_t_CEP_psi}
\end{center}
\end{figure}

Let us first discuss pair production in the case of linearly polarized colliding pulses. The differential particle distributions in longitudinal $z$-coordinate, calculated by means of Eqs.~(\ref{w_e_average1}) -- (\ref{H_b_single}) for the values $0$, $\pi/2$, $\pi$ of relative phase $\Psi$ and for CEP $\tilde{\varphi}=0,\pi/4,\pi/2$, are shown in Fig.~\ref{dN_chi_CEP_psi}. The distributions possess a spiky structure, with the peaks positions sensitive to $\Psi$ but independent of $\tilde{\varphi}$. This feature is obvious from the form of the simplified expression (\ref{epsilon_mu-1_lin1}) of the invariant $\epsilon$ in electric regime. Furthermore, the production rate is maximal for $\Psi = 0$ and $\tilde{\varphi} = \pi/2$. As $\Psi$ varies, the peaks are shifted from the focus center (as is the case e.g., for $\Psi = \pi/2$ or $\pi$), hence the distribution becomes asymmetric (changes from unimodal to bimodal) and the production rate is reduced. For $\Psi=\pi$ the bimodal distribution becomes symmetric. In contrast to the position, the separation of the peaks is merely independent of phase shifts (remains about $\pi\Delta^2=0.0314$ in dimensionless units used at the figure). The same is approximately true also for the peak widths. It is seen that for the adopted values of parameters pair production results the generation of narrow (FWHM = $0.0636\mu m$) particle-antiparticle bunches localized in longitudinal direction.

Similar features are observed also in a temporal distribution (see Fig.~\ref{dN_t_CEP_psi}), where, however, the distribution profiles (peak heights as well as their locations) are sensitive to both phase shifts. In all the cases shown, in agreement with Eq.~(\ref{epsilon_mu-1_lin1}), the temporal distribution is unimodal and maximal for $\tilde{\varphi}+\Psi/2\approx \pi/2$ and bimodal symmetric for $\tilde{\varphi}+\Psi/2\approx 0$ or $\pi$. The peaks FWHM width is here as narrow as $200$as. 
As expected from Fig.~\ref{epsilon_contour_lin}, the most prolific production rate is observed in Fig.~\ref{dN_t_CEP_psi}(c) for $\tilde{\varphi}~=~\pi/2$ and $\Psi = 0$, i.e., for an in-phase configuration of the counterpropagating beams. 
 
\subsection{Differential pair production rate (circular polarization)}

For colliding circularly e-polarized laser pulses it has been observed earlier \cite{ChitradipCEP} that the structure of the invariant electric and magnetic fields and the pair production rate depend on their relative handedness. In particular, in RR configuration the invariant fields and the differential particle production rates do not reveal any CEP dependence. When the counterpropagating pulses are in-phase, a broad unimodal temporal particle distribution is produced. On the other hand, the colliding pulses in RL configuration produce particle bunches localized in time and with notable CEP dependence. It is, therefore, natural to consider these two cases separately.
 
 \begin{figure}[t]
  \begin{center}
  \includegraphics[width = 3.5in]{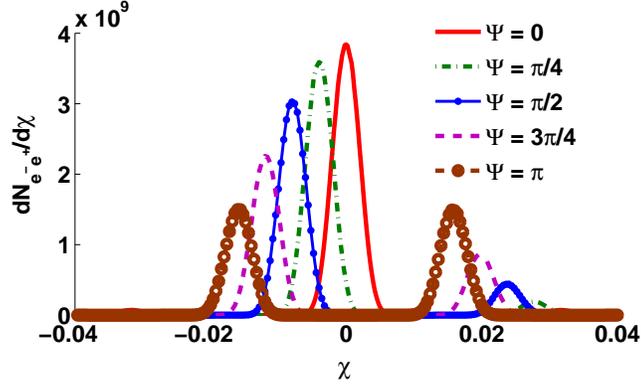} 
  \caption{Spatial longitudinal distributions of $e^+e^-$ pairs created in RR configuration for $\Psi=0,\,\pi/4,\,\pi/2,\,3\pi/4,\,\pi$. The laser parameters are same as in Fig.~\ref{epsilon_contour_lin}.}
  \label{dN_chi_cir_CEP0_RL_Phase}
  \end{center}
  \end{figure}
For RR configuration, since the invariant fields trivially depend on time and are independent on $\tilde{\varphi}$ [see Eq.~(\ref{epsilon_eta_RR_cir}) and note that the same is true exactly], it is enough to present the spatial distribution of created pairs only in dependence on $\Psi$, see Fig.~\ref{dN_chi_cir_CEP0_RL_Phase}. The maximal production rate is achieved with $\Psi = 0$, in this case the distribution looks unimodal and symmetric. The minimal production rate corresponds to a symmetric bimodal distribution at $\Psi = \pi$, while for the intermediate values of $\Psi$ the distribution is bimodal but asymmetric. The FWHM width (about $0.0764\mu m$) of the peaks, as well as separation between them ($\lambda/2=0.5\mu m$), are both insensitive to the phase $\Psi$. All these results are in obvious agreement with our above discussion of the invariant field structure [see Eq.~(\ref{epsilon_eta_RR_cir}) and Fig.~\ref{Epsilon_chi_RR_phase_cir}].

\begin{figure}[t]
 \begin{center}
 \includegraphics[width = 5.5in]{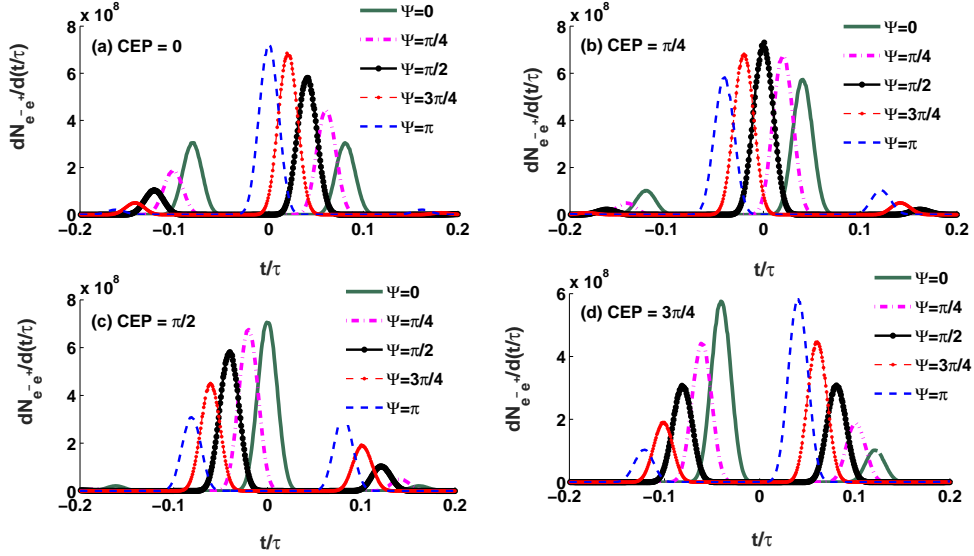} 
 \caption{Evolution of the $e^+e^-$ pair production rate  in RL configuration for  $\tilde{\varphi} =  0,~\pi/4,~\pi/2,~3\pi/4$ and for $\Psi =~0,~\pi/4,~\pi/2,~3\pi/4,~\pi$.  Laser parameters are the same as in Fig.~\ref{epsilon_contour_lin}.}
 \label{dN_t_cir_CEP_RL_Phase}
 \end{center}
 \end{figure}

 
For RL configuration, in contrast, the invariant fields trivially depend on position, but are sensitive to both phase shifts, hence it is enough to present only temporal evolution of the production rate, but in dependence on both phase shifts, see  Fig.~\ref{dN_t_cir_CEP_RL_Phase}.
As in previous cases, at variation of phases the distribution profile changes from unimodal through asymmetric bimodal to symmetric bimodal. In agreement with Eq.~(\ref{epsilon_eta_cir_CEP}), the particular form of the distribution depends solely on the combination $\tilde{\varphi}+\Psi/2$, with maximal and minimal production rates when it is close to $\pi/2$ and to zero or $\pi$, respectively. The FWHM width of the generated particle bunches is about $220as$, much shorter than laser pulse duration $10fs$. As before, it is insensitive to both phases $\tilde{\varphi}$ and $\Psi$, as well as to separation between the peaks when the profiles are bimodal, which is about $1.6fs$.

By comparing the figures of this section, we conclude that for the same values of parameters the circularly polarized RR-configuration with $\Psi=0$ maximizes the total number of created pairs. Namely this configuration was discussed in greater details (in particular, in dependence of amplitude and focusing degree) in Ref.~\cite{bula}.
  
\section{conclusion}\label{conclusion}

It was proposed \cite{bula,PhysRevLett.104.220404,PhysRevD.82.105026Hebenstreit,PhysRevE.71.016404,PhysRevE.69.036408,PhysRevD.91.125026Wollert,PhysRevA.86.2012Gonoskov} that coherent superposition of focused optical laser pulses is favorable for future observations of spontaneous pair production below the Schwinger limit because of constructive interference, which notably increases the peak field strength. In this context, it is natural to analyze phase dependence of the arising interference pattern, as well as the possibilities for phase control of the corresponding pair production rate. Here we have done it with respect to both phase shifts inherent to the problem, the carrier envelope phase $\tilde{\varphi}$ of individual pulses, and their relative phase shift $\Psi$. As shown, their variation shifts the spiky interference pattern of the invariant electric field with respect to the carrier envelope, leaving the widths of the peaks unaltered. 

The same conclusion broadly refers to the pair production rate, which depends on the invariant fields monotonously. 
Indeed, in all the cases considered here we could relate the peculiarities of particle distribution to the underlying invariant field structure.
However, since due to exponentiation in Eqs.~(\ref{w_e_average}) or (\ref{w_e_average1}), at the level of production rate only those peaks of the invariant electric field that are closest to the center of the spatiotemporal envelope remain significant, the resulting spatial and temporal distributions of created pairs can look either nearly unimodal or bimodal. For the same reason, their longitudinal and temporal spike widths are much smaller than of the original invariant field structure, meaning time-localized (during hundreds attoseconds) formation of extremely short (tens nanometers long) electron-positron bunches. The post production dynamics of such dense bunches may be highly non-trivial \cite{PhysRevE.69.036408} and may need a separate study.

We predict that the total pair production is maximal when one of the spikes is located near the center of the spatiotemporal envelope (and distribution of created pairs looks approximately unimodal), and minimal when the neighboring spikes are off-centered but located symmetrically. The particular phase shifts required for each case depend on polarization of the pulses. Among the considered cases, for the parameters adopted here the global maximum is achieved with a circularly polarized RR configuration considered in Ref.~\cite{bula}. 

Possibility of phase control of Schwinger pair production under discussion may be useful, e.g., to increase the attainable intensity of tightly focused colliding laser pulses by reducing pair production and hence preventing field depletion at their crossing, or, conversely, to measure the typically unknown field structure and phase relations of extremely strong laser pulses in a way similar to proposed in Ref.~\cite{PhysRevLett.105.063903} by using the multiphoton Compton scattering.

\begin{acknowledgments}

A. M. F. acknowledges support from the MEPhI Academic Excellence	Project (Contract No.	02.a03.21.0005), Tomsk State University Competitiveness Improvement Program, and the Russian Fund for Basic Research (Grant 16-02-00963a).
\end{acknowledgments}

\appendix
\section{Simplified temporal envelope for counterpropagating pulses in a focal region}\label{Appendix}


As shown in Ref.~\cite{Narozhny2000}, finite pulse duration can be incorporated into a focused pulse model roughly by introducing an individual envelope factor $g(\varphi)$  for each pulse [such that $g(0) = 1$ and $g(\varphi)$ vanishes for $|\varphi|\gtrsim\omega\tau$, where $\tau$ is pulse duration]. Let us show that inside a focal region of counterpropagating pulses one can with high accuracy rather use a single common envelope instead. Consider their total (for definiteness, electric) field 
 \begin{equation}
  \textbf{E}_{tot} = g_1 \textbf{E}_f + g_2 \textbf{E}_b ,
 \end{equation}
where $ \textbf{E}_{f,b} $ are the fields of forward and backward propagating pulses [see Eqs.~(\ref{E_f_single}), (\ref{E_b_single})],
$g_1 = g_1(\varphi/{\omega\tau}) = g_1((t-z/c)/\tau)$ and $g_2 = g_2(\varphi^{\prime}/{\omega\tau}) = g_2((t+z/c)/\tau)$ -- their individual envelopes. For the sake of simplicity, we assume $g_1(\varphi) =g_2(\varphi)=\exp(-4\varphi^2/\tau^2)$, as used throughout the paper. By rewriting
  \begin{equation}
   \textbf{E}_{tot} = \sqrt{g_1g_2}\Big(\sqrt{g_1/g_2} \textbf{E}_f + \sqrt{g_2/g_1} \textbf{E}_b \Big),
  \end{equation}
 we define $g = \sqrt{g_1g_2}= \exp(-4t^2/\tau^2-4z^2/{c^2\tau^2})$, then the weight factors $g_f = \sqrt{g_1/g_2}$ and $g_b = \sqrt{g_2/g_1}$ read 
  $g_{f,b} = \exp(\pm 8zt/{c\tau^2})= \exp(\pm 8\chi t^{\prime}L/{c\tau})$, where $\chi = z/L$ and $t^{\prime} = t/\tau$ are the dimensionless longitudinal coordinate and  time normalized by laser duration. Even though for the parameters used here ($\tau = 10fs, \lambda = 1\mu m , \Delta = 0.1$) we have $L/{c\tau} = 5.3$, pair creation is localized in a tiny region $|\chi|\lesssim 0.02$ and $|t'|\lesssim 0.15$ (see Figs.~\ref{dN_chi_CEP_psi}-\ref{dN_t_cir_CEP_RL_Phase}), where both weight factors $g_{f,b}$ are close to unity. Hence one can optionally use for the total field a modified common envelope function $g = \sqrt{g_1g_2}$, which is extremely useful to explain the results in a qualitative way. The numerical results are in good agreement in both cases of using either $g_1$ and $g_2$ or the approximate common envelope $g$. 
\bibliographystyle{apsrev}
\bibliography{ref}

\begin{thebibliography}{55}
\expandafter\ifx\csname natexlab\endcsname\relax\def\natexlab#1{#1}\fi
\expandafter\ifx\csname bibnamefont\endcsname\relax
  \def\bibnamefont#1{#1}\fi
\expandafter\ifx\csname bibfnamefont\endcsname\relax
  \def\bibfnamefont#1{#1}\fi
\expandafter\ifx\csname citenamefont\endcsname\relax
  \def\citenamefont#1{#1}\fi
\expandafter\ifx\csname url\endcsname\relax
  \def\url#1{\texttt{#1}}\fi
\expandafter\ifx\csname urlprefix\endcsname\relax\def\urlprefix{URL }\fi
\providecommand{\bibinfo}[2]{#2}
\providecommand{\eprint}[2][]{\url{#2}}

\bibitem[{\citenamefont{Berestetskii et~al.}(2012)\citenamefont{Berestetskii,
  Pitaevskii, and Lifshitz}}]{berestetskii2012quantum}
\bibinfo{author}{\bibfnamefont{V.}~\bibnamefont{Berestetskii}},
  \bibinfo{author}{\bibfnamefont{L.}~\bibnamefont{Pitaevskii}},
  \bibnamefont{and} \bibinfo{author}{\bibfnamefont{E.}~\bibnamefont{Lifshitz}},
  \emph{\bibinfo{title}{Quantum Electrodynamics}}, \bibinfo{number}{Vol. 4}
  (\bibinfo{publisher}{Elsevier Science}, \bibinfo{year}{2012}).

\bibitem[{\citenamefont{Dirac}(1930)}]{dirac1930theory}
\bibinfo{author}{\bibfnamefont{P.~A.} \bibnamefont{Dirac}}, in
  \emph{\bibinfo{booktitle}{Proceedings of the Royal Society of London A:
  Mathematical, Physical and Engineering Sciences}} (\bibinfo{organization}{The
  Royal Society}, \bibinfo{year}{1930}), vol. \bibinfo{volume}{126}, pp.
  \bibinfo{pages}{360--365}.

\bibitem[{\citenamefont{Ritus and Nikishov}(1979)}]{viritus}
\bibinfo{author}{\bibfnamefont{V.~I.} \bibnamefont{Ritus}} \bibnamefont{and}
  \bibinfo{author}{\bibfnamefont{A.~I.} \bibnamefont{Nikishov}},
  \emph{\bibinfo{title}{''Quantum Electrodynamics of phenomena in a Strong
  Field''}}, Trudy Fiz. Inst. Akad. Nauk SSSR, No. 111
  (\bibinfo{publisher}{Moscow}, \bibinfo{year}{1979}).

\bibitem[{\citenamefont{Di~Piazza et~al.}(2012)\citenamefont{Di~Piazza,
  M\"uller, Hatsagortsyan, and Keitel}}]{RevModPhys.84.1177DiPiazza}
\bibinfo{author}{\bibfnamefont{A.}~\bibnamefont{Di~Piazza}},
  \bibinfo{author}{\bibfnamefont{C.}~\bibnamefont{M\"uller}},
  \bibinfo{author}{\bibfnamefont{K.~Z.} \bibnamefont{Hatsagortsyan}},
  \bibnamefont{and} \bibinfo{author}{\bibfnamefont{C.~H.}
  \bibnamefont{Keitel}}, \bibinfo{journal}{Rev. Mod. Phys.}
  \textbf{\bibinfo{volume}{84}}, \bibinfo{pages}{1177} (\bibinfo{year}{2012}).

\bibitem[{\citenamefont{Sauter}(1931)}]{Sauter}
\bibinfo{author}{\bibfnamefont{F.}~\bibnamefont{Sauter}}, \bibinfo{journal}{Z.
  Phys.} \textbf{\bibinfo{volume}{69}}, \bibinfo{pages}{742}
  (\bibinfo{year}{1931}).

\bibitem[{\citenamefont{Heisenberg and Euler}(1936)}]{Heisenberg}
\bibinfo{author}{\bibfnamefont{W.}~\bibnamefont{Heisenberg}} \bibnamefont{and}
  \bibinfo{author}{\bibfnamefont{H.}~\bibnamefont{Euler}}, \bibinfo{journal}{Z.
  Phys.} \textbf{\bibinfo{volume}{98}}, \bibinfo{pages}{714}
  (\bibinfo{year}{1936}).

\bibitem[{\citenamefont{Schwinger}(1951)}]{PhysRev.82.664}
\bibinfo{author}{\bibfnamefont{J.}~\bibnamefont{Schwinger}},
  \bibinfo{journal}{Phys. Rev.} \textbf{\bibinfo{volume}{82}},
  \bibinfo{pages}{664} (\bibinfo{year}{1951}).

\bibitem[{\citenamefont{Volkov}(1935)}]{VolkovDM}
\bibinfo{author}{\bibfnamefont{D.}~\bibnamefont{Volkov}},
  \bibinfo{journal}{Zeitschrift f{\"u}r Physik} \textbf{\bibinfo{volume}{94}},
  \bibinfo{pages}{250} (\bibinfo{year}{1935}).

\bibitem[{\citenamefont{Narozhnyi and Nikishov}(1970)}]{NarozhnyiNikishov1970}
\bibinfo{author}{\bibfnamefont{N.}~\bibnamefont{Narozhnyi}} \bibnamefont{and}
  \bibinfo{author}{\bibfnamefont{A.}~\bibnamefont{Nikishov}},
  \bibinfo{journal}{Yadern. Fiz. 11: 1072-7.}  (\bibinfo{year}{1970}).

\bibitem[{\citenamefont{Nikishov}(1970{\natexlab{a}})}]{Nikishov1970346}
\bibinfo{author}{\bibfnamefont{A.}~\bibnamefont{Nikishov}},
  \bibinfo{journal}{Nuclear Physics B} \textbf{\bibinfo{volume}{21}},
  \bibinfo{pages}{346 } (\bibinfo{year}{1970}{\natexlab{a}}).

\bibitem[{\citenamefont{Brezin and Itzykson}(1970)}]{PhysRevD.2.1191Itzykson}
\bibinfo{author}{\bibfnamefont{E.}~\bibnamefont{Brezin}} \bibnamefont{and}
  \bibinfo{author}{\bibfnamefont{C.}~\bibnamefont{Itzykson}},
  \bibinfo{journal}{Phys. Rev. D} \textbf{\bibinfo{volume}{2}},
  \bibinfo{pages}{1191} (\bibinfo{year}{1970}).

\bibitem[{\citenamefont{Allor et~al.}(2008)\citenamefont{Allor, Cohen, and
  McGady}}]{SchwingerGraphene2008}
\bibinfo{author}{\bibfnamefont{D.}~\bibnamefont{Allor}},
  \bibinfo{author}{\bibfnamefont{T.~D.} \bibnamefont{Cohen}}, \bibnamefont{and}
  \bibinfo{author}{\bibfnamefont{D.~A.} \bibnamefont{McGady}},
  \bibinfo{journal}{Phys. Rev. D} \textbf{\bibinfo{volume}{78}},
  \bibinfo{pages}{096009} (\bibinfo{year}{2008}).

\bibitem[{\citenamefont{Fillion-Gourdeau and
  MacLean}(2015)}]{PhysRevBGrapheneSchw2015}
\bibinfo{author}{\bibfnamefont{F.}~\bibnamefont{Fillion-Gourdeau}}
  \bibnamefont{and} \bibinfo{author}{\bibfnamefont{S.}~\bibnamefont{MacLean}},
  \bibinfo{journal}{Phys. Rev. B} \textbf{\bibinfo{volume}{92}},
  \bibinfo{pages}{035401} (\bibinfo{year}{2015}).

\bibitem[{\citenamefont{Fillion-Gourdeau
  et~al.}(2016)\citenamefont{Fillion-Gourdeau, Gagnon, Lefebvre, and
  MacLean}}]{PhysRevBGrapheneSchw2016}
\bibinfo{author}{\bibfnamefont{F.}~\bibnamefont{Fillion-Gourdeau}},
  \bibinfo{author}{\bibfnamefont{D.}~\bibnamefont{Gagnon}},
  \bibinfo{author}{\bibfnamefont{C.}~\bibnamefont{Lefebvre}}, \bibnamefont{and}
  \bibinfo{author}{\bibfnamefont{S.}~\bibnamefont{MacLean}},
  \bibinfo{journal}{Phys. Rev. B} \textbf{\bibinfo{volume}{94}},
  \bibinfo{pages}{125423} (\bibinfo{year}{2016}).

\bibitem[{\citenamefont{Mourou et~al.}(2006)\citenamefont{Mourou, Tajima, and
  Bulanov}}]{RevModPhys.78.309Mourou2006}
\bibinfo{author}{\bibfnamefont{G.~A.} \bibnamefont{Mourou}},
  \bibinfo{author}{\bibfnamefont{T.}~\bibnamefont{Tajima}}, \bibnamefont{and}
  \bibinfo{author}{\bibfnamefont{S.~V.} \bibnamefont{Bulanov}},
  \bibinfo{journal}{Rev. Mod. Phys.} \textbf{\bibinfo{volume}{78}},
  \bibinfo{pages}{309} (\bibinfo{year}{2006}).

\bibitem[{\citenamefont{Yanovsky et~al.}(2008)\citenamefont{Yanovsky, Chvykov,
  Kalinchenko, Rousseau, Planchon, Matsuoka, Maksimchuk, Nees, Cheriaux, Mourou
  et~al.}}]{Yanovsky:08}
\bibinfo{author}{\bibfnamefont{V.}~\bibnamefont{Yanovsky}},
  \bibinfo{author}{\bibfnamefont{V.}~\bibnamefont{Chvykov}},
  \bibinfo{author}{\bibfnamefont{G.}~\bibnamefont{Kalinchenko}},
  \bibinfo{author}{\bibfnamefont{P.}~\bibnamefont{Rousseau}},
  \bibinfo{author}{\bibfnamefont{T.}~\bibnamefont{Planchon}},
  \bibinfo{author}{\bibfnamefont{T.}~\bibnamefont{Matsuoka}},
  \bibinfo{author}{\bibfnamefont{A.}~\bibnamefont{Maksimchuk}},
  \bibinfo{author}{\bibfnamefont{J.}~\bibnamefont{Nees}},
  \bibinfo{author}{\bibfnamefont{G.}~\bibnamefont{Cheriaux}},
  \bibinfo{author}{\bibfnamefont{G.}~\bibnamefont{Mourou}},
  \bibnamefont{et~al.}, \bibinfo{journal}{Opt. Express}
  \textbf{\bibinfo{volume}{16}}, \bibinfo{pages}{2109} (\bibinfo{year}{2008}).

\bibitem[{\citenamefont{Tajima and Mourou}(2002)}]{PhysRevSTAB.5.031301}
\bibinfo{author}{\bibfnamefont{T.}~\bibnamefont{Tajima}} \bibnamefont{and}
  \bibinfo{author}{\bibfnamefont{G.}~\bibnamefont{Mourou}},
  \bibinfo{journal}{Phys. Rev. ST Accel. Beams} \textbf{\bibinfo{volume}{5}},
  \bibinfo{pages}{031301} (\bibinfo{year}{2002}).

\bibitem[{\citenamefont{Burke et~al.}(1997)\citenamefont{Burke, Field,
  Horton-Smith, Spencer, Walz, Berridge, Bugg, Shmakov, Weidemann, Bula
  et~al.}}]{DbrukePhysRevLett.79.1626}
\bibinfo{author}{\bibfnamefont{D.~L.} \bibnamefont{Burke}},
  \bibinfo{author}{\bibfnamefont{R.~C.} \bibnamefont{Field}},
  \bibinfo{author}{\bibfnamefont{G.}~\bibnamefont{Horton-Smith}},
  \bibinfo{author}{\bibfnamefont{J.~E.} \bibnamefont{Spencer}},
  \bibinfo{author}{\bibfnamefont{D.}~\bibnamefont{Walz}},
  \bibinfo{author}{\bibfnamefont{S.~C.} \bibnamefont{Berridge}},
  \bibinfo{author}{\bibfnamefont{W.~M.} \bibnamefont{Bugg}},
  \bibinfo{author}{\bibfnamefont{K.}~\bibnamefont{Shmakov}},
  \bibinfo{author}{\bibfnamefont{A.~W.} \bibnamefont{Weidemann}},
  \bibinfo{author}{\bibfnamefont{C.}~\bibnamefont{Bula}}, \bibnamefont{et~al.},
  \bibinfo{journal}{Phys. Rev. Lett.} \textbf{\bibinfo{volume}{79}},
  \bibinfo{pages}{1626} (\bibinfo{year}{1997}).

\bibitem[{\citenamefont{Bell and Kirk}(2008)}]{PhysRevLett.101.200403Bell}
\bibinfo{author}{\bibfnamefont{A.~R.} \bibnamefont{Bell}} \bibnamefont{and}
  \bibinfo{author}{\bibfnamefont{J.~G.} \bibnamefont{Kirk}},
  \bibinfo{journal}{Phys. Rev. Lett.} \textbf{\bibinfo{volume}{101}},
  \bibinfo{pages}{200403} (\bibinfo{year}{2008}).

\bibitem[{\citenamefont{Nerush et~al.}(2011)\citenamefont{Nerush, Kostyukov,
  Fedotov, Narozhny, Elkina, and Ruhl}}]{PhysRevLett.106.035001Nerush}
\bibinfo{author}{\bibfnamefont{E.~N.} \bibnamefont{Nerush}},
  \bibinfo{author}{\bibfnamefont{I.~Y.} \bibnamefont{Kostyukov}},
  \bibinfo{author}{\bibfnamefont{A.~M.} \bibnamefont{Fedotov}},
  \bibinfo{author}{\bibfnamefont{N.~B.} \bibnamefont{Narozhny}},
  \bibinfo{author}{\bibfnamefont{N.~V.} \bibnamefont{Elkina}},
  \bibnamefont{and} \bibinfo{author}{\bibfnamefont{H.}~\bibnamefont{Ruhl}},
  \bibinfo{journal}{Phys. Rev. Lett.} \textbf{\bibinfo{volume}{106}},
  \bibinfo{pages}{035001} (\bibinfo{year}{2011}).

\bibitem[{\citenamefont{Bula et~al.}(1996)\citenamefont{Bula, McDonald, Prebys,
  Bamber, Boege, Kotseroglou, Melissinos, Meyerhofer, Ragg, Burke
  et~al.}}]{CbulaPhysRevLett.76.3116}
\bibinfo{author}{\bibfnamefont{C.}~\bibnamefont{Bula}},
  \bibinfo{author}{\bibfnamefont{K.~T.} \bibnamefont{McDonald}},
  \bibinfo{author}{\bibfnamefont{E.~J.} \bibnamefont{Prebys}},
  \bibinfo{author}{\bibfnamefont{C.}~\bibnamefont{Bamber}},
  \bibinfo{author}{\bibfnamefont{S.}~\bibnamefont{Boege}},
  \bibinfo{author}{\bibfnamefont{T.}~\bibnamefont{Kotseroglou}},
  \bibinfo{author}{\bibfnamefont{A.~C.} \bibnamefont{Melissinos}},
  \bibinfo{author}{\bibfnamefont{D.~D.} \bibnamefont{Meyerhofer}},
  \bibinfo{author}{\bibfnamefont{W.}~\bibnamefont{Ragg}},
  \bibinfo{author}{\bibfnamefont{D.~L.} \bibnamefont{Burke}},
  \bibnamefont{et~al.}, \bibinfo{journal}{Phys. Rev. Lett.}
  \textbf{\bibinfo{volume}{76}}, \bibinfo{pages}{3116} (\bibinfo{year}{1996}).

\bibitem[{\citenamefont{Narozhny and Fofanov}(2000)}]{Narozhny2000}
\bibinfo{author}{\bibfnamefont{N.}~\bibnamefont{Narozhny}} \bibnamefont{and}
  \bibinfo{author}{\bibfnamefont{M.}~\bibnamefont{Fofanov}},
  \bibinfo{journal}{JETP} \textbf{\bibinfo{volume}{90}}, \bibinfo{pages}{753}
  (\bibinfo{year}{2000}).

\bibitem[{\citenamefont{Fedotov}(2009)}]{fedotov2009electron}
\bibinfo{author}{\bibfnamefont{A.}~\bibnamefont{Fedotov}},
  \bibinfo{journal}{Laser physics} \textbf{\bibinfo{volume}{19}},
  \bibinfo{pages}{214} (\bibinfo{year}{2009}).

\bibitem[{\citenamefont{Salamin et~al.}(2002)\citenamefont{Salamin, Mocken, and
  Keitel}}]{PhysRevSTAB_salamin}
\bibinfo{author}{\bibfnamefont{Y.~I.} \bibnamefont{Salamin}},
  \bibinfo{author}{\bibfnamefont{G.~R.} \bibnamefont{Mocken}},
  \bibnamefont{and} \bibinfo{author}{\bibfnamefont{C.~H.}
  \bibnamefont{Keitel}}, \bibinfo{journal}{Phys. Rev. ST Accel. Beams}
  \textbf{\bibinfo{volume}{5}}, \bibinfo{pages}{101301} (\bibinfo{year}{2002}).

\bibitem[{\citenamefont{Su et~al.}(2012{\natexlab{a}})\citenamefont{Su, Jiang,
  Lv, Li, Sheng, Grobe, and Su}}]{PhysRevA2012Su}
\bibinfo{author}{\bibfnamefont{W.}~\bibnamefont{Su}},
  \bibinfo{author}{\bibfnamefont{M.}~\bibnamefont{Jiang}},
  \bibinfo{author}{\bibfnamefont{Z.~Q.} \bibnamefont{Lv}},
  \bibinfo{author}{\bibfnamefont{Y.~J.} \bibnamefont{Li}},
  \bibinfo{author}{\bibfnamefont{Z.~M.} \bibnamefont{Sheng}},
  \bibinfo{author}{\bibfnamefont{R.}~\bibnamefont{Grobe}}, \bibnamefont{and}
  \bibinfo{author}{\bibfnamefont{Q.}~\bibnamefont{Su}}, \bibinfo{journal}{Phys.
  Rev. A} \textbf{\bibinfo{volume}{86}}, \bibinfo{pages}{013422}
  (\bibinfo{year}{2012}{\natexlab{a}}).

\bibitem[{\citenamefont{Su et~al.}(2012{\natexlab{b}})\citenamefont{Su, Su, Lv,
  Jiang, Lu, Sheng, and Grobe}}]{PhysRevLett.109.253202Su}
\bibinfo{author}{\bibfnamefont{Q.}~\bibnamefont{Su}},
  \bibinfo{author}{\bibfnamefont{W.}~\bibnamefont{Su}},
  \bibinfo{author}{\bibfnamefont{Q.~Z.} \bibnamefont{Lv}},
  \bibinfo{author}{\bibfnamefont{M.}~\bibnamefont{Jiang}},
  \bibinfo{author}{\bibfnamefont{X.}~\bibnamefont{Lu}},
  \bibinfo{author}{\bibfnamefont{Z.~M.} \bibnamefont{Sheng}}, \bibnamefont{and}
  \bibinfo{author}{\bibfnamefont{R.}~\bibnamefont{Grobe}},
  \bibinfo{journal}{Phys. Rev. Lett.} \textbf{\bibinfo{volume}{109}},
  \bibinfo{pages}{253202} (\bibinfo{year}{2012}{\natexlab{b}}).

\bibitem[{\citenamefont{Bulanov et~al.}(2006)\citenamefont{Bulanov, Narozhny,
  Mur, and Popov}}]{bula}
\bibinfo{author}{\bibfnamefont{S.}~\bibnamefont{Bulanov}},
  \bibinfo{author}{\bibfnamefont{N.}~\bibnamefont{Narozhny}},
  \bibinfo{author}{\bibfnamefont{V.}~\bibnamefont{Mur}}, \bibnamefont{and}
  \bibinfo{author}{\bibfnamefont{V.}~\bibnamefont{Popov}},
  \bibinfo{journal}{JETP} \textbf{\bibinfo{volume}{102}}, \bibinfo{pages}{9}
  (\bibinfo{year}{2006}).

\bibitem[{\citenamefont{Bulanov
  et~al.}(2010{\natexlab{a}})\citenamefont{Bulanov, Mur, Narozhny, Nees, and
  Popov}}]{PhysRevLett.104.220404}
\bibinfo{author}{\bibfnamefont{S.~S.} \bibnamefont{Bulanov}},
  \bibinfo{author}{\bibfnamefont{V.~D.} \bibnamefont{Mur}},
  \bibinfo{author}{\bibfnamefont{N.~B.} \bibnamefont{Narozhny}},
  \bibinfo{author}{\bibfnamefont{J.}~\bibnamefont{Nees}}, \bibnamefont{and}
  \bibinfo{author}{\bibfnamefont{V.~S.} \bibnamefont{Popov}},
  \bibinfo{journal}{Phys. Rev. Lett.} \textbf{\bibinfo{volume}{104}},
  \bibinfo{pages}{220404} (\bibinfo{year}{2010}{\natexlab{a}}).

\bibitem[{\citenamefont{Hebenstreit et~al.}(2010)\citenamefont{Hebenstreit,
  Alkofer, and Gies}}]{PhysRevD.82.105026Hebenstreit}
\bibinfo{author}{\bibfnamefont{F.}~\bibnamefont{Hebenstreit}},
  \bibinfo{author}{\bibfnamefont{R.}~\bibnamefont{Alkofer}}, \bibnamefont{and}
  \bibinfo{author}{\bibfnamefont{H.}~\bibnamefont{Gies}},
  \bibinfo{journal}{Phys. Rev. D} \textbf{\bibinfo{volume}{82}},
  \bibinfo{pages}{105026} (\bibinfo{year}{2010}).

\bibitem[{\citenamefont{Bulanov et~al.}(2005)\citenamefont{Bulanov, Fedotov,
  and Pegoraro}}]{PhysRevE.71.016404}
\bibinfo{author}{\bibfnamefont{S.~S.} \bibnamefont{Bulanov}},
  \bibinfo{author}{\bibfnamefont{A.~M.} \bibnamefont{Fedotov}},
  \bibnamefont{and} \bibinfo{author}{\bibfnamefont{F.}~\bibnamefont{Pegoraro}},
  \bibinfo{journal}{Phys. Rev. E} \textbf{\bibinfo{volume}{71}},
  \bibinfo{pages}{016404} (\bibinfo{year}{2005}).

\bibitem[{\citenamefont{Bulanov}(2004)}]{PhysRevE.69.036408}
\bibinfo{author}{\bibfnamefont{S.~S.} \bibnamefont{Bulanov}},
  \bibinfo{journal}{Phys. Rev. E} \textbf{\bibinfo{volume}{69}},
  \bibinfo{pages}{036408} (\bibinfo{year}{2004}).

\bibitem[{\citenamefont{W\"ollert et~al.}(2015)\citenamefont{W\"ollert, Bauke,
  and Keitel}}]{PhysRevD.91.125026Wollert}
\bibinfo{author}{\bibfnamefont{A.}~\bibnamefont{W\"ollert}},
  \bibinfo{author}{\bibfnamefont{H.}~\bibnamefont{Bauke}}, \bibnamefont{and}
  \bibinfo{author}{\bibfnamefont{C.~H.} \bibnamefont{Keitel}},
  \bibinfo{journal}{Phys. Rev. D} \textbf{\bibinfo{volume}{91}},
  \bibinfo{pages}{125026} (\bibinfo{year}{2015}).

\bibitem[{\citenamefont{Gonoskov et~al.}(2012)\citenamefont{Gonoskov, Aiello,
  Heugel, and Leuchs}}]{PhysRevA.86.2012Gonoskov}
\bibinfo{author}{\bibfnamefont{I.}~\bibnamefont{Gonoskov}},
  \bibinfo{author}{\bibfnamefont{A.}~\bibnamefont{Aiello}},
  \bibinfo{author}{\bibfnamefont{S.}~\bibnamefont{Heugel}}, \bibnamefont{and}
  \bibinfo{author}{\bibfnamefont{G.}~\bibnamefont{Leuchs}},
  \bibinfo{journal}{Phys. Rev. A} \textbf{\bibinfo{volume}{86}},
  \bibinfo{pages}{053836} (\bibinfo{year}{2012}).

\bibitem[{\citenamefont{Kim and Page}(2002)}]{PhysRevDSPKim2002}
\bibinfo{author}{\bibfnamefont{S.~P.} \bibnamefont{Kim}} \bibnamefont{and}
  \bibinfo{author}{\bibfnamefont{D.~N.} \bibnamefont{Page}},
  \bibinfo{journal}{Phys. Rev. D} \textbf{\bibinfo{volume}{65}},
  \bibinfo{pages}{105002} (\bibinfo{year}{2002}).

\bibitem[{\citenamefont{Kim and Lee}(2007)}]{PhysRevDSPKim2007}
\bibinfo{author}{\bibfnamefont{S.~P.} \bibnamefont{Kim}} \bibnamefont{and}
  \bibinfo{author}{\bibfnamefont{H.~K.} \bibnamefont{Lee}},
  \bibinfo{journal}{Phys. Rev. D} \textbf{\bibinfo{volume}{76}},
  \bibinfo{pages}{125002} (\bibinfo{year}{2007}).

\bibitem[{\citenamefont{Kim and Page}(2006)}]{PhysRevD.73SPKim2006}
\bibinfo{author}{\bibfnamefont{S.~P.} \bibnamefont{Kim}} \bibnamefont{and}
  \bibinfo{author}{\bibfnamefont{D.~N.} \bibnamefont{Page}},
  \bibinfo{journal}{Phys. Rev. D} \textbf{\bibinfo{volume}{73}},
  \bibinfo{pages}{065020} (\bibinfo{year}{2006}).

\bibitem[{\citenamefont{Dunne}(2009)}]{Dunne2009}
\bibinfo{author}{\bibfnamefont{G.~V.} \bibnamefont{Dunne}},
  \bibinfo{journal}{The European Physical Journal D}
  \textbf{\bibinfo{volume}{55}}, \bibinfo{pages}{327} (\bibinfo{year}{2009}).

\bibitem[{\citenamefont{Kim et~al.}(2008)\citenamefont{Kim, Lee, and
  Yoon}}]{PhysRevDSPKim2008}
\bibinfo{author}{\bibfnamefont{S.~P.} \bibnamefont{Kim}},
  \bibinfo{author}{\bibfnamefont{H.~K.} \bibnamefont{Lee}}, \bibnamefont{and}
  \bibinfo{author}{\bibfnamefont{Y.}~\bibnamefont{Yoon}},
  \bibinfo{journal}{Phys. Rev. D} \textbf{\bibinfo{volume}{78}},
  \bibinfo{pages}{105013} (\bibinfo{year}{2008}).

\bibitem[{\citenamefont{Hwang and Kim}(2009)}]{PhysRevDSPKim2009}
\bibinfo{author}{\bibfnamefont{W.-Y.~P.} \bibnamefont{Hwang}} \bibnamefont{and}
  \bibinfo{author}{\bibfnamefont{S.~P.} \bibnamefont{Kim}},
  \bibinfo{journal}{Phys. Rev. D} \textbf{\bibinfo{volume}{80}},
  \bibinfo{pages}{065004} (\bibinfo{year}{2009}).

\bibitem[{\citenamefont{Hebenstreit and
  Fillion-Gourdeau}(2014)}]{Hebenstreit2014189}
\bibinfo{author}{\bibfnamefont{F.}~\bibnamefont{Hebenstreit}} \bibnamefont{and}
  \bibinfo{author}{\bibfnamefont{F.}~\bibnamefont{Fillion-Gourdeau}},
  \bibinfo{journal}{Physics Letters B} \textbf{\bibinfo{volume}{739}},
  \bibinfo{pages}{189 } (\bibinfo{year}{2014}).

\bibitem[{\citenamefont{Chervyakov and
  Kleinert}(2009)}]{PhysRevD.80Chervyakov2009}
\bibinfo{author}{\bibfnamefont{A.}~\bibnamefont{Chervyakov}} \bibnamefont{and}
  \bibinfo{author}{\bibfnamefont{H.}~\bibnamefont{Kleinert}},
  \bibinfo{journal}{Phys. Rev. D} \textbf{\bibinfo{volume}{80}},
  \bibinfo{pages}{065010} (\bibinfo{year}{2009}).

\bibitem[{\citenamefont{Gies and Klingm\"uller}(2005)}]{PhysRevDHGies2005}
\bibinfo{author}{\bibfnamefont{H.}~\bibnamefont{Gies}} \bibnamefont{and}
  \bibinfo{author}{\bibfnamefont{K.}~\bibnamefont{Klingm\"uller}},
  \bibinfo{journal}{Phys. Rev. D} \textbf{\bibinfo{volume}{72}},
  \bibinfo{pages}{065001} (\bibinfo{year}{2005}).

\bibitem[{\citenamefont{Kleinert et~al.}(2008)\citenamefont{Kleinert, Ruffini,
  and Xue}}]{PhysRevDHKleinert2008}
\bibinfo{author}{\bibfnamefont{H.}~\bibnamefont{Kleinert}},
  \bibinfo{author}{\bibfnamefont{R.}~\bibnamefont{Ruffini}}, \bibnamefont{and}
  \bibinfo{author}{\bibfnamefont{S.-S.} \bibnamefont{Xue}},
  \bibinfo{journal}{Phys. Rev. D} \textbf{\bibinfo{volume}{78}},
  \bibinfo{pages}{025011} (\bibinfo{year}{2008}).

\bibitem[{\citenamefont{Abdukerim et~al.}(2013)\citenamefont{Abdukerim, Li, and
  Xie}}]{AbdukerimCarrier}
\bibinfo{author}{\bibfnamefont{N.}~\bibnamefont{Abdukerim}},
  \bibinfo{author}{\bibfnamefont{Z.-L.} \bibnamefont{Li}}, \bibnamefont{and}
  \bibinfo{author}{\bibfnamefont{B.-S.} \bibnamefont{Xie}},
  \bibinfo{journal}{Physics Letters B} \textbf{\bibinfo{volume}{726}},
  \bibinfo{pages}{820 } (\bibinfo{year}{2013}).

\bibitem[{\citenamefont{Hebenstreit et~al.}(2009)\citenamefont{Hebenstreit,
  Alkofer, Dunne, and Gies}}]{PhysRevLett_CEP2009}
\bibinfo{author}{\bibfnamefont{F.}~\bibnamefont{Hebenstreit}},
  \bibinfo{author}{\bibfnamefont{R.}~\bibnamefont{Alkofer}},
  \bibinfo{author}{\bibfnamefont{G.~V.} \bibnamefont{Dunne}}, \bibnamefont{and}
  \bibinfo{author}{\bibfnamefont{H.}~\bibnamefont{Gies}},
  \bibinfo{journal}{Phys. Rev. Lett.} \textbf{\bibinfo{volume}{102}},
  \bibinfo{pages}{150404} (\bibinfo{year}{2009}).

\bibitem[{\citenamefont{Dumlu and Dunne}(2010)}]{PhysRevLettDumluStokes}
\bibinfo{author}{\bibfnamefont{C.~K.} \bibnamefont{Dumlu}} \bibnamefont{and}
  \bibinfo{author}{\bibfnamefont{G.~V.} \bibnamefont{Dunne}},
  \bibinfo{journal}{Phys. Rev. Lett.} \textbf{\bibinfo{volume}{104}},
  \bibinfo{pages}{250402} (\bibinfo{year}{2010}).

\bibitem[{\citenamefont{Gies and Torgrimsson}(2016)}]{PhysRevLettGiesHolger}
\bibinfo{author}{\bibfnamefont{H.}~\bibnamefont{Gies}} \bibnamefont{and}
  \bibinfo{author}{\bibfnamefont{G.}~\bibnamefont{Torgrimsson}},
  \bibinfo{journal}{Phys. Rev. Lett.} \textbf{\bibinfo{volume}{116}},
  \bibinfo{pages}{090406} (\bibinfo{year}{2016}).

\bibitem[{\citenamefont{Dumlu}(2010)}]{PhysRevD.82Dumlu}
\bibinfo{author}{\bibfnamefont{C.~K.} \bibnamefont{Dumlu}},
  \bibinfo{journal}{Phys. Rev. D} \textbf{\bibinfo{volume}{82}},
  \bibinfo{pages}{045007} (\bibinfo{year}{2010}).

\bibitem[{\citenamefont{Kohlf\"urst et~al.}(2013)\citenamefont{Kohlf\"urst,
  Mitter, von Winckel, Hebenstreit, and Alkofer}}]{PhysRevD.88.045028Kohlfurst}
\bibinfo{author}{\bibfnamefont{C.}~\bibnamefont{Kohlf\"urst}},
  \bibinfo{author}{\bibfnamefont{M.}~\bibnamefont{Mitter}},
  \bibinfo{author}{\bibfnamefont{G.}~\bibnamefont{von Winckel}},
  \bibinfo{author}{\bibfnamefont{F.}~\bibnamefont{Hebenstreit}},
  \bibnamefont{and} \bibinfo{author}{\bibfnamefont{R.}~\bibnamefont{Alkofer}},
  \bibinfo{journal}{Phys. Rev. D} \textbf{\bibinfo{volume}{88}},
  \bibinfo{pages}{045028} (\bibinfo{year}{2013}).

\bibitem[{\citenamefont{Banerjee and Singh}(2017)}]{Banerjee2017}
\bibinfo{author}{\bibfnamefont{C.}~\bibnamefont{Banerjee}} \bibnamefont{and}
  \bibinfo{author}{\bibfnamefont{M.~P.} \bibnamefont{Singh}},
  \bibinfo{journal}{JETP} \textbf{\bibinfo{volume}{125}}, \bibinfo{pages}{12}
  (\bibinfo{year}{2017}).

\bibitem[{\citenamefont{Banerjee and Singh}(2018)}]{ChitradipCEP}
\bibinfo{author}{\bibfnamefont{C.}~\bibnamefont{Banerjee}} \bibnamefont{and}
  \bibinfo{author}{\bibfnamefont{M.~P.} \bibnamefont{Singh}},
  \bibinfo{journal}{Eur. Phys. J. D} \textbf{\bibinfo{volume}{72}},
  \bibinfo{pages}{4} (\bibinfo{year}{2018}).

\bibitem[{\citenamefont{Bulanov
  et~al.}(2010{\natexlab{b}})\citenamefont{Bulanov, Mur, Narozhny, Nees, and
  Popov}}]{PhysRevLettBulanov}
\bibinfo{author}{\bibfnamefont{S.~S.} \bibnamefont{Bulanov}},
  \bibinfo{author}{\bibfnamefont{V.~D.} \bibnamefont{Mur}},
  \bibinfo{author}{\bibfnamefont{N.~B.} \bibnamefont{Narozhny}},
  \bibinfo{author}{\bibfnamefont{J.}~\bibnamefont{Nees}}, \bibnamefont{and}
  \bibinfo{author}{\bibfnamefont{V.~S.} \bibnamefont{Popov}},
  \bibinfo{journal}{Phys. Rev. Lett.} \textbf{\bibinfo{volume}{104}},
  \bibinfo{pages}{220404} (\bibinfo{year}{2010}{\natexlab{b}}).

\bibitem[{\citenamefont{Nikishov}(1970{\natexlab{b}})}]{nikishov1970pair}
\bibinfo{author}{\bibfnamefont{A.}~\bibnamefont{Nikishov}},
  \bibinfo{journal}{Sov. Phys. JETP} \textbf{\bibinfo{volume}{30}},
  \bibinfo{pages}{660} (\bibinfo{year}{1970}{\natexlab{b}}).

\bibitem[{\citenamefont{Bashmakov et~al.}(2014)\citenamefont{Bashmakov, Nerush,
  Kostyukov, Fedotov, and Narozhny}}]{ElectricregimeFedotov}
\bibinfo{author}{\bibfnamefont{V.~F.} \bibnamefont{Bashmakov}},
  \bibinfo{author}{\bibfnamefont{E.~N.} \bibnamefont{Nerush}},
  \bibinfo{author}{\bibfnamefont{I.~Y.} \bibnamefont{Kostyukov}},
  \bibinfo{author}{\bibfnamefont{A.~M.} \bibnamefont{Fedotov}},
  \bibnamefont{and} \bibinfo{author}{\bibfnamefont{N.~B.}
  \bibnamefont{Narozhny}}, \bibinfo{journal}{Physics of Plasmas}
  \textbf{\bibinfo{volume}{21}}, \bibinfo{pages}{013105}
  (\bibinfo{year}{2014}).

\bibitem[{\citenamefont{Mackenroth et~al.}(2010)\citenamefont{Mackenroth,
  Di~Piazza, and Keitel}}]{PhysRevLett.105.063903}
\bibinfo{author}{\bibfnamefont{F.}~\bibnamefont{Mackenroth}},
  \bibinfo{author}{\bibfnamefont{A.}~\bibnamefont{Di~Piazza}},
  \bibnamefont{and} \bibinfo{author}{\bibfnamefont{C.~H.}
  \bibnamefont{Keitel}}, \bibinfo{journal}{Phys. Rev. Lett.}
  \textbf{\bibinfo{volume}{105}}, \bibinfo{pages}{063903}
  (\bibinfo{year}{2010}).

\end{thebibliography}
\end{document}